\documentclass[preprint,pre,aps,amsfonts,amsmath,superscriptaddress,nofootinbib,showkeys]{revtex4}
\usepackage{amsfonts,amsmath,amssymb,amscd}
\usepackage{mathrsfs}
\usepackage{empheq}
\usepackage{graphicx}
\usepackage{bm,bbold}
\usepackage{color}
\usepackage{calligra}
\usepackage[T1]{fontenc}
\begin{document}
%
%
\title{Schr\"{o}dinger Equation Driven by the Square of a Gaussian Field: Instanton Analysis in the Large Amplification Limit}
\author{Philippe Mounaix}
\email{philippe.mounaix@polytechnique.edu}
\affiliation{CPHT, CNRS, \'Ecole
polytechnique, Institut Polytechnique de Paris, 91120 Palaiseau, France.}
\date{\today}
\begin{abstract}
We study the tail of $p(U)$, the probability distribution of $U=\vert\psi(0,L)\vert^2$, for $\ln U\gg 1$, $\psi(x,z)$ being the solution to $\partial_z\psi -\frac{i}{2m}\nabla_{\perp}^2 \psi =g\vert S\vert^2\, \psi$, where $S(x,z)$ is a complex Gaussian random field, $z$ and $x$ respectively are the axial and transverse coordinates, with $0\le z\le L$, and both $m\ne 0$ and $g>0$ are real parameters. We perform the first instanton analysis of the corresponding Martin-Siggia-Rose action, from which it is found that the realizations of $S$ concentrate onto long filamentary instantons, as $\ln U\to +\infty$. The tail of $p(U)$ is deduced from the statistics of the instantons. The value of $g$ above which $\langle U\rangle$ diverges coincides with the one obtained by the completely different approach developed in Mounaix et al. 2006 {\it Commun. Math. Phys.} {\bf 264}~741. Numerical simulations clearly show a statistical bias of $S$ towards the instanton for the largest sampled values of $\ln U$. The high maxima --- or `hot spots' --- of $\vert S(x,z)\vert^2$ for the biased realizations of $S$ tend to cluster in the instanton region.
\end{abstract}
\maketitle
%
%
\section{Introduction}\label{intro}
In the second part of their seminal paper on the breakdown of linear instability in stimulated Brillouin scattering\ \cite{RD1994}, Rose and DuBois investigated the following equation for the complex amplitude $\psi(x,z)$ of the scattered light electric field
\begin{equation}\label{withDeq}
\left\lbrace
\begin{array}{l}
\partial_z\psi(x,z)-\frac{i}{2m}\nabla_{\perp}^2 \psi(x,z)=g\vert S(x,z)\vert^2\psi(x,z), \\
0\le z\le L,\ x\in\Lambda\subset\mathbb{R}^d,\ {\rm and}\ \psi(x,0)=1.
\end{array}\right.
\end{equation}
In Eq.\ (\ref{withDeq}), $z$ and $x$ respectively denote the axial and transverse coordinates in a plasma of length $L$ and cross-sectional domain $\Lambda$ (often a torus like, e.g., in mathematics oriented work and/or numerical simulations using spectral methods). The boundary condition at $z=0$ is taken to be a constant for simplicity and $m\ne 0$ is a real parameter introduced for convenience. In Ref.~\cite{RD1994}, the coupling constant $g>0$ is proportional to the average laser intensity and the complex amplitude of the laser electric field $S(x,z)$ is a homogeneous Gaussian random field with zero mean and normalized intensity $\langle\vert S(x,z)\vert^2\rangle =1$. For our purposes, we can be less restrictive and take $S(x,z)$ transversally homogeneous with normalization $L^{-1}\int_0^L \langle\vert S(x,z)\vert^2\rangle\, dz =1$. From now on, we accept the idealizations inherent in the derivation of Eq.\ (\ref{withDeq}), setting aside the question of its validity as a realistic model, which varies from one physical problem to the other. As a stochastic PDE, the diffraction-amplification problem\ (\ref{withDeq}) is a Schr\"{o}dinger equation driven by the square of a Gaussian field.

Using heuristic arguments and numerical simulations, Rose and DuBois found that the expected value of the scattered energy density, $\langle\vert\psi(x_0,L)\vert^2\rangle$, at some given $x_0\in\Lambda$ diverges for every $L>0$ when $g$ is greater than some critical value, $g_c(L)$, yet to be determined. Here, the average $\langle\vert\psi\vert^2\rangle$ is taken over the realizations of the Gaussian field $S$. Physically, this divergence was interpreted in\ \cite{RD1994} as indicating the breakdown of the linear model (\ref{withDeq}) and the onset of a saturated nonlinear regime in high overintensities, or hot spots, of $\vert S(x,z)\vert^2$. We will shortly come back to the role of the hot spots in the divergence of $\langle\vert\psi(x_0,L)\vert^2\rangle$. Note that in the limit referred to in\ \cite{RD1994} as the independent hot spot model, this divergence was pointed out by Akhmanov {\it et al.} 20 years before\ \cite{ADP1974}. The problem was then analyzed in\ \cite{ADLM2001,ML2004,MCL2006} from a more rigorous mathematical point of view, establishing the numerical results of Ref.~\cite{RD1994} on much firmer ground and giving the exact expression of the critical coupling $g_c(L)$. In the following, we will take $x_0=0$ without loss of generality (by statistical invariance under $x$-translation) and we will write $U=\vert\psi(0,L)\vert^2$.

Whether or not $\langle U\rangle$ diverges depends on the extreme upper tail of $p(U)$ --- the probability distribution function (PDF) of $U$ --- in the limit of a large $\ln U$ (see the beginning of Sec.~\ref{withoutDsec}). It is then natural to ask what the realizations of $S(x,z)$ yielding a large $\ln U$ are like, with $U$ and $S(x,z)$ related to each other through Eq.~(\ref{withDeq}), from which probability distribution they are drawn, and if the corresponding tail of $p(U)$ does give the correct value of $g_c(L)$. Answering these questions is the subject of this paper.

To put our work into perspective, it is interesting to recall how the existence of $g_c(L)$ has been interpreted in laser-plasma physics literature since Ref.\ \cite{RD1994}. The interpretation relies on the implicit assumption that the realizations of $\vert S(x,z)\vert^2$ giving rise to a large $\ln U\gg\langle\ln U\rangle$ and the generic ones for which $\ln U\simeq\langle\ln U\rangle$ are alike, in the sense of being made up of local, statistically independent, overintensities, or hot spots, separated from each other by a few correlation lengths of $S(x,z)$\ \cite{Dixit1993,G1985,RD1993}. Hot spot contribution to the amplification of $\vert\psi\vert^2$ can then be computed by using the remarkable result that intense hot spots have a non-random profile depending on the correlation function of $S(x,z)$ and being the same for each hot spot\ \cite{RD1993,A1981}. Thus, intense hot spots are entirely characterized by their random intensity which turns out to be exponentially distributed (for large intensity and to within slow, algebraic, corrections)\ \cite{RD1993,G1999}. For $g$ large enough, intense hot spots become statistically significant as the exponentially large amplification they produce outbalances their exponentially small scarcity, leading to the divergence of $\langle U\rangle$. The smallest value of $g$ at which this divergence  occurs defines the critical coupling $g_c(L)$ and for $g>g_c(L)$ physics could be expected to be dominated by intense hot spots. Unfortunately, this interpretation fails to give the correct value of $g_c(L)$ for $L$ greater than a hot spot length\ \cite{Mounaix2001,MD2004}. The assumption of high intensity, statistically independent hot spots giving the dominant contribution to the amplification in the large $\ln U$ limit must be revisited. Large values of $\ln U$ are produced by rare realizations of $S(x,z)$ that have no reason {\it a priori} to look like generic realizations with no other structures than uncorrelated, local hot spots randomly scattered in $\Lambda\times\lbrack 0,L\rbrack$. It may or may not be so: the answer will come out of the calculations.

In the simpler diffraction-free case where $m^{-1}=0$ in Eq.~(\ref{withDeq}), the problem reduces to a mere $1D$ amplification along $z$ with $\psi(L)=\exp\left(g\int_0^L \vert S(z)\vert^2 dz\right)$. A large value of $\ln\vert\psi(L)\vert^2$ corresponds to a large value of $\int_0^L \vert S(z)\vert^2 dz$. Thus, the realizations of $S(z)$ that form the tail of $p(U)$ are the ones with a large $L^2$-norm. These realizations were studied thoroughly in\ \cite{MD2004,MC2011,MMB2012}. Let $C(z,z^\prime)=\langle S(z)S(z^\prime)^\ast \rangle$ and define the covariance operator $T_C$ acting on $f(z)\in L^2([0,L])$ by
\begin{equation}\label{covariance1D}
(T_C f)(z) =\int_0^L C(z,z^\prime)\, f(z^\prime)\, dz^\prime ,
\end{equation}
with $0\le z\le L$. As a correlation function, $C$ is a positive definite kernel and all the eigenvalues of $T_C$ are necessarily real and positive. Write $\mu_1>0$ the largest eigenvalue of $T_C$ with degeneracy $d_1$. It was proved in\ \cite{MD2004,MC2011} that the realizations of $S(z)$ with a large $L^2$-norm concentrate onto the fundamental eigenspace of $T_C$, i.e., the eigenspace associated with the largest eigenvalue $\mu_1$. More specifically, writing $\lbrace \phi_1,\cdots ,\phi_{d_1}\rbrace$ an orthonormal basis of the fundamental eigenspace of $T_C$, one has
\begin{equation}\label{oldresultwithoutD1}
S(z)\sim\sqrt{\eta} \sum_{i=\nu}^{d_1} a_\nu \phi_\nu(z)\ \ \ \ \ (\| S\|_2\to +\infty),
\end{equation}
with $\eta\sim\| S\|_2^2$, where $\|\cdot\|_2$ denotes the $L^2$-norm over $\lbrack 0,L\rbrack$. The $a_\nu$s are complex numbers normalized to $\sum_{\nu=1}^{d_1}\vert a_\nu\vert^2 =1$. The probability distribution of $\eta$ has the gamma-distribution tail $p(\eta)\sim\eta^{d_1-1}{\rm e}^{-\eta/\mu_1}$ for large $\eta$, and the $a_\nu$s define a random $2d_1$-dimensional (real) unit vector $\bm{a}$ with coordinates ${\rm Re}(a_\nu)$ and ${\rm Im}(a_\nu)$ ($1\le\nu\le d_1$) the direction of which is uniformly distributed over the unit $(2d_1 -1)$-sphere. From Eq.~(\ref{oldresultwithoutD1}) it is clear that the realizations of $S(z)$ with a large $L^2$-norm are less random than the Gaussian field $S(z)$ itself. It only takes $2d_1$ random quantities to characterize these realizations entirely: $\eta$ and the direction of $\bm{a}$. For instance, if $\mu_1$ is not degenerate ($d_1=1$), Eq.\ (\ref{oldresultwithoutD1}) yields
\begin{equation}\label{oldresultwithoutD2}
\frac{S(z)}{\| S\|_2}\sim {\rm e}^{i\theta} \phi_1(z)\ \ \ \ \ (\| S\|_2\to +\infty),
\end{equation}
where $\theta$ is a random phase uniformly distributed over $\lbrack 0,2\pi)$ and $\vert S(z)\vert/\| S\|_2\sim\vert\phi_1(z)\vert$ is non-random, which means that the profile of $S(z)$ is purely deterministic in this case. Note that Eq.~(\ref{oldresultwithoutD2}) rules out any description in terms of localized hot spots when $L$ is large, as $\phi_1(z)$ typically is a one-bump delocalized mode spreading over the whole domain $0\le z\le L$ (see\ \cite{MD2004} for details). As will be seen further on, the randomness reduction of $S$ when the amplification is large occurs in the $m^{-1}\ne 0$ case too.

From $U=\vert\psi(L)\vert^2=\exp(2g\| S\|_2^2)$ and $\eta\sim\| S\|_2^2$ as $\| S\|_2 \to +\infty$, one gets $\eta\sim (2g)^{-1}\ln U$ as $\ln U\to +\infty$. The tail of $p(U)$ is then readily obtained from $p(\eta)\sim\eta^{d_1-1}{\rm e}^{-\eta/\mu_1}$ and the change of variables from $\eta$ to $U$. One finds, in logarithmic form,
\begin{eqnarray}\label{tailpofUwithoutD}
\ln p(U)&\sim&\ln\left\lbrack\frac{1}{U}\, p\left(\eta=\frac{1}{2g}\ln U\right)\right\rbrack \\
&=& -\left(1+\frac{1}{2\mu_1g}\right)\, \ln U +\left( d_1 -1\right)\, \ln\ln U +O(1) \nonumber
\ \ \ \ \ (\ln U\to +\infty),
\end{eqnarray}
from which it follows that $p(U)$ has a leading algebraic tail $\propto U^{-\zeta}$ (modulated by logarithmic corrections in the amplitude) with exponent $\zeta =(1+1/2\mu_1 g)$ depending continuously on the parameters of the model. Injecting this result into $\langle U\rangle =\int_{1}^{+\infty}Up(U)\, dU$, one finds that the critical coupling in the diffraction-free case is given by $g_c(L)=1/2\mu_1$ (where $\mu_1$ depends on $L$)\ \cite{MD2004}. Note that it is also possible to determine the tail of $p(U)$ exactly from the full Gaussian statistics of $S$, which makes it possible to estimate the contribution of the subleading corrections to Eq.~(\ref{oldresultwithoutD1}). Skipping the details, one finds that these corrections do not contribute to $\ln p(U)$ by terms greater than $O(1)$ as $\ln U\to +\infty$.

To conclude this brief overview of diffraction-free results, let us mention the interesting connection between the concentration onto the fundamental eigenspace of $T_C$ in Eq.\ (\ref{oldresultwithoutD1}) and the Bose-Einstein condensation of $S(z)$ in the `thermodynamic' limit defined by $L\to +\infty$ and $\| S\|_2^2\to +\infty$ with fixed $\| S\|_2^2/L$, see\ \cite{MMB2012}. Note also that more general concentration properties can be found in the limit where the large $L^2$-norm is replaced with a large quadratic or linear form, see\ \cite{Mounaix2015,Mounaix2019}.

By contrast, much less is known in the general case with diffraction where $m^{-1}\ne 0$ in Eq.~(\ref{withDeq}). The only results so far are the numerical ones in the second part of\ \cite{RD1994} and the analytical calculation of the critical coupling performed in\ \cite{MCL2006}, where it is proved that the critical coupling without diffraction cannot be less than the one with diffraction, the latter being given by
\begin{equation}\label{critcouplingwithD}
g_c(L)=\frac{1}{2\sup_{x(\cdot)\in B(0,L)}\mu_1\lbrack x(\cdot)\rbrack},
\end{equation}
and the former by $1/2\mu_1\lbrack x(\cdot)\equiv 0\rbrack$. In Eq.~(\ref{critcouplingwithD}), $B(0,L)$ denotes the set of all the continuous paths in $\Lambda$ satisfying $x(L)=0$ and $\mu_1\lbrack x(\cdot)\rbrack$ is the largest eigenvalue of the covariance operator $T_{x(\cdot)}$ defined by Eq.~(\ref{covariance1D}) with $C(z,z^\prime)=\langle S(x(z),z)S(x(z^\prime),z^\prime)^\ast \rangle$.

The question then arises whether the non-local quantity $\sup_{x(\cdot)\in B(0,L)}\mu_1\lbrack x(\cdot)\rbrack$ in the expression for $g_c(L)$ is the signature of a corresponding non-local structure in the realizations of $S(x,z)$ giving rise to a large $\ln U$. (Note that only local quantities computed on the hot spot scale can appear in the hot spot model.) To answer this question we need to find a way to identify such realizations. The corresponding tail of $p(U)$ will then be tested in return by checking that the critical coupling it yields coincides with the one in Eq.~(\ref{critcouplingwithD}). The calculations in\ \cite{MD2004,MC2011} are of no help as being specific to the diffraction-free case. To determine $S(x,z)$ when $\ln U$ is large and get the tail of $p(U)$ in the presence of diffraction we need a different approach.

A possible line of attack is through the functional integral formalism introduced by Janssen\ \cite{Janssen1976}, DeDominicis\ \cite{DeDominicis1976,DDP1978}, and Phythian\ \cite{Phythian1977} (see also\ \cite{JP1979,Jensen1981}). This method provides a formal description of classical statistical dynamics in terms of functional integrals analogous to Feynman's action-integral formalism of quantum theory. Applying the method to the stochastic equation\ (\ref{withDeq}), one finds that $p(U)$ can be formally written as the functional integral
\begin{equation}\label{generalactionint}
p(U)=\int_{\varphi(x,0)=1}
\delta(U-\vert\varphi(0,L)\vert^2)\, {\rm e}^{\mathcal{A}}\, 
\mathscr{D}^2\varphi\, \mathscr{D}^2\tilde{\varphi}\, \mathscr{D}^2 S,
\end{equation}
where $\varphi$ and $\tilde{\varphi}$ are complex Martin-Siggia-Rose conjugate fields\ \cite{MSR1973}, $\mathscr{D}^2\equiv\mathscr{D}{\rm Re}(\cdot)\mathscr{D}{\rm Im}(\cdot)$, and $\mathcal{A}\equiv\mathcal{A}(\varphi,\tilde{\varphi},S)$ is an `action' depending on $\varphi$, $\tilde{\varphi}$, and $S$, often referred to as the Martin-Siggia-Rose (MSR) action in the literature. When $\ln U$ is large, the functional integral on the right-hand side of Eq.\ (\ref{generalactionint}) is determined by the field configuration --- called the `leading instanton' --- corresponding to the highest saddle-point of the action $\mathcal{A}$. Note that, usually, $S$ is integrated out in Eq.\ (\ref{generalactionint}), leading to
\begin{equation}\label{actionintwithoutS}
p(U)=\int_{\varphi(x,0)=1}
\delta(U-\vert\varphi(0,L)\vert^2)\, {\rm e}^{\mathcal{L}}\, 
\mathscr{D}^2\varphi\, \mathscr{D}^2\tilde{\varphi},
\end{equation}
where the action $\mathcal{L}\equiv\mathcal{L}(\varphi,\tilde{\varphi})$ is defined by
\begin{equation}\label{actionL}
{\rm e}^{\mathcal{L}(\varphi,\tilde{\varphi})}
=\int {\rm e}^{\mathcal{A}(\varphi,\tilde{\varphi},S)}\,\mathscr{D}^2 S.
\end{equation}
The expressions of $p(U)$ in Eqs.\ (\ref{generalactionint}) and\ (\ref{actionintwithoutS}) are equivalent. Either can be used and the large $\ln U$ behavior of $p(U)$ can equally be obtained from an instanton analysis of Eq.~(\ref{actionintwithoutS}), instead of Eq.~(\ref{generalactionint}). As we want to determine the most probable realizations of $S(x,z)$ when $\ln U$ is large, it is natural to keep $S$ explicit and to use the functional representation\ (\ref{generalactionint}).

At large $\ln U$ and fixed instanton, the fluctuations of the fields around the instanton are small and can be integrated out in Eq.~(\ref{generalactionint}) as standard Gaussian fluctuations, yielding the leading asymptotic behavior
\begin{equation}\label{tailpofUgeneral}
\ln p(U)\sim\ln\left\langle\delta(U-\vert\varphi_{\rm inst}(0,L)\vert^2)\right\rangle_{S_{\rm inst}}
\ \ \ \ \ (\ln U\to +\infty),
\end{equation}
where the subscript `inst' stands for leading instanton and $\langle\cdot\rangle_{S_{\rm inst}}$ denotes the average over the realizations of $S_{\rm inst}$. In the diffraction-free case, it will be checked in Sec.\ \ref{withoutDsec} that $S_{\rm inst}$ coincides with the right-hand side of Eq.\ (\ref{oldresultwithoutD1}) and that the tail of $p(U)$ in Eq.~(\ref{tailpofUgeneral}) is the same as the one in Eq.\ (\ref{tailpofUwithoutD}).

One of the early attempts to obtain PDF tails from the highest saddle-point of the action in a functional integral representation was made by Giles for Navier-Stokes turbulence\ \cite{Giles1995}. Unfortunately, the perturbative approach followed in this work was doomed to fail as instantons are nonperturbative objects by nature. The nonperturbative instanton analysis of intermittency in fluid turbulence was initiated by Falkovich {\it et al.} in\ \cite{FKLM1996}, but many questions are still open\ \cite{SGMG2022,AMPV2022}. Since then, advances and new applications of the instanton approach in stochastic field theories have been made. A relatively recent review on fluid turbulence applications can be found in\ \cite{GGS2015}. Different kinds of nonlinear Schr\"{o}dinger equations with additive noise have been analyzed using instanton calculus within the last twenty years, see e.g.\ \cite{FKLT2001}, among others. See also the instanton analyses of PDF tails in forced Burgers turbulence in\ \cite{GM1996,BM1996} and the numerical results in\ \cite{GGS2013}. Instanton solutions have been tested numerically for the Kardar-Parisi-Zhang equation in\ \cite{HMS2019}. Here, we give the first instanton analysis of the stochastic amplifier~(\ref{withDeq}) in the large amplification limit of interest in the overcritical regime $g>g_c(L)$. Our strategy is two-step:

\begin{itemize}
\item[$\bullet$]{write the MSR action $\mathcal{A}(\varphi,\tilde{\varphi},S)$ for the stochastic amplifier\ (\ref{withDeq}) and find the corresponding leading instanton $S_{\rm{inst}}$. The realizations of $S_{\rm{inst}}$ define the driver onto which the realizations of $S$ concentrate in the large $\ln U$ limit, which generalizes the diffraction-free result in Eq.\ (\ref{oldresultwithoutD1}) to the case with diffraction;}
\item[$\bullet$]{use the instanton $\varphi_{\rm{inst}}$ and $S_{\rm{inst}}$ obtained at the first step on the right-hand side of Eq.\ (\ref{tailpofUgeneral}) to get the tail of $p(U)$.}
\end{itemize}

Before entering the details of the calculations, it is useful to give a brief summary of the main new results obtained in this paper.

\begin{itemize}
\item[$\diamond$]{We show that in the large $\ln U$ limit, the realizations of $S$ concentrate onto large-scale filamentary instantons running along specific non-random paths in $B(0,L)$. Each of these paths, denoted by $x_{\rm inst}(\cdot)$, is a path maximizing the largest eigenvalue $\mu_1\lbrack x(\cdot)\rbrack$ of the covariance operator $T_{x(\cdot)}$ defined in Eq.~(\ref{covariance3Dpath}). In the case of a `single-filament instanton' (see Sec.~\ref{singlepathsec}) and assuming a non-degenerate $\mu_1\lbrack x_{\rm inst}(\cdot)\rbrack$, we prove that}
\end{itemize}
\begin{equation}\label{summaryintro1}
S(x,z)\sim\frac{c_1}{\mu_{\rm max}}
\, \int_0^L C(x-x_{\rm inst}(z^\prime),z,z^\prime)
\, \phi_1(z^\prime)\, dz^\prime\ \ \ \ \ (\ln U\to +\infty),
\end{equation}

\begin{itemize}
\item[]{where $C(x-x^\prime ,z,z^\prime)=\langle S(x,z)S(x^\prime ,z^\prime)^\ast \rangle$, $\mu_{\rm max}=\mu_1\lbrack x_{\rm inst}(\cdot)\rbrack$ with normalized eigenfunction $\phi_1$, and $c_1$ is a complex Gaussian random variable with $\langle c_1\rangle =\langle c_1^2\rangle =0$ and $\langle\vert c_1\vert^2\rangle =\mu_{\rm max}$. The instanton on the right-hand side of Eq.~(\ref{summaryintro1}) lives within a long thin tube, or filament, running along $x_{\rm inst}(\cdot)$ (see the end of Sec.~\ref{singlepathLI}). Normalizing $S$ to its $L^2$-norm, Eq.~(\ref{summaryintro1}) yields}
\end{itemize}
\begin{equation*}
\frac{S(x,z)}{\| S\|_2}\sim A{\rm e}^{i\arg(c_1)} \int_0^L C(x-x_{\rm inst}(z^\prime),z,z^\prime)
\, \phi_1(z^\prime)\, dz^\prime\ \ \ \ \ (\ln U\to +\infty),
\end{equation*}

\begin{itemize}
\item[]{where $A>0$ is a constant, and the profile of $S(x,z)$ defined by $\vert S(x,z)\vert/\| S\|_2$ is asymptotically non-random as $\ln U\to +\infty$.}
\end{itemize}

\begin{itemize}
\item[$\diamond$]{We determine the tail of $p(U)$ for large $\ln U$ from the statistics of the instanton on the right-hand side of Eq.~(\ref{summaryintro1}). We find that $p(U)$ has a leading algebraic tail $\propto U^{-\zeta}$ with exponent $\zeta =(1+1/2\mu_{\rm max} g)$, modulated by a slow varying amplitude (slower than algebraic). Injecting this result into $\langle U\rangle =\int_{1}^{+\infty}Up(U)\, dU$, we find that $\langle U\rangle$ diverges for all $g>1/2\mu_{\rm max}$. The critical coupling is thus given by $g_c(L)=1/2\mu_{\rm max}$, where $\mu_{\rm max}$ depends on $L$, in agreement with Eq.~(\ref{critcouplingwithD}). We can then explain the intriguing presence of the non-local quantity $\mu_{\rm max}$ in the expression of $g_c(L)$ as a direct consequence of the fact that the realizations of $S$ causing the divergence of $\langle U\rangle$ are realizations of the non-local instanton (\ref{summaryintro1}), rather than of localized hot spots, as is widely assumed.}
\item[$\diamond$]{Finally, the emergence of the instanton in the realizations of $S$ as $\ln U$ increases is observed in numerical simulations as a statistical bias of $S$ towards the instanton. For the largest sampled values of $\ln U$, the emerging large-scale instanton coexists with small-scale hot spots in a non-negligible fraction of realizations. The presence of the instanton causes the hot spots to cluster in the instanton region instead of being uniformly scattered in $\Lambda\times\lbrack 0,L\rbrack$, and the level of $\vert S(x,z)\vert^2$ between the hot spots remains significantly higher than it would be in the absence of instanton.}
\end{itemize}

The paper is organized as follows. In Section~\ref{withoutDsec}, we test the functional approach by revisiting the diffraction-free problem where the results are already known. In Section~\ref{withDsec}, we write the instanton equations for the full problem with diffraction in the case of one transverse dimension ($d=1$) and we specify the class of $S$ we consider. Section~\ref{singlepathsec} is devoted to the solution of the instanton equations in the case of `single-filament' instantons. The corresponding tail of $p(U)$ is determined. In Section~\ref{numerics}, we report on numerical simulations for realizations of $S$ in a sample of experimentally realistic size. Finally, we discuss our results and their implications, especially in laser-matter interaction physics, and we give potential perspectives in Section~\ref{conclusion}. Some technical material is relegated to the appendices.
%
%
\section{Amplification without diffraction revisited}\label{withoutDsec}
As a warm-up to the full problem~(\ref{withDeq}), we test the functional approach on the simpler problem without diffraction and see how the results in Eqs.\ (\ref{oldresultwithoutD1}) and\ (\ref{tailpofUwithoutD}) can also be obtained from an instanton analysis of the appropriate MSR action.

Before we start, it is useful to briefly come back to the definition of the asymptotic limit. In the absence of diffraction, $U=\exp\left(2g\| S\|_2^2\right)$ and the limit $\| S\|_2^2\to +\infty$ in\ \cite{MD2004,MC2011} reads $\ln U\to +\infty$, which defines the asymptotic limit (rather than $U\to +\infty$). The same applies to the case with diffraction, where $U$ is also the result of exponential amplification.
%
%
\subsection{The MSR action $\bm{\mathcal{A}(\varphi,\tilde{\varphi},S)}$}\label{actionwithoutDsec}
In the diffraction-free limit, $m^{-1}=0$, the equation\ (\ref{withDeq}) reduces to the $1D$ stochastic amplifier (for fixed $x$, not written)
\begin{equation}\label{withoutDeq}
\left\lbrace
\begin{array}{l}
d_z\psi(z)-g\vert S(z)\vert^2\psi(z)=0, \\
0\le z\le L\ {\rm and}\ \psi(0)=1.
\end{array}\right.
\end{equation}
Let $F\lbrack\psi(\cdot)\rbrack$ be a functional of $\psi(z)$ solution to Eq.\ (\ref{withoutDeq}). From the general formalism developed in\ \cite{Janssen1976,DeDominicis1976,DDP1978,Phythian1977,JP1979,Jensen1981} it can be shown that $F\lbrack\psi(\cdot)\rbrack$ admits the functional integral representation
\begin{equation}\label{FunctionalwithoutD}
F\lbrack\psi(\cdot)\rbrack =\int_{\varphi(0)=1} F\lbrack\varphi(\cdot)\rbrack\, 
{\rm e}^{\frac{i}{2}\, \left(\left\langle\tilde{\varphi}\left\vert d_z -g\vert S\vert^2\right\vert\varphi\right\rangle
+c.\, c.\right)}\, \mathscr{D}^2\varphi\, \mathscr{D}^2\tilde{\varphi},
\end{equation}
with Dirac's bracket notation $\langle f\vert O\vert h\rangle=\int_0^L f(z)^\ast (Oh)(z)\, dz$. Note that since $\psi(0)$ is real, $\psi(z)$ is also real for all $z$ and a representation with real $\varphi$ and $\tilde{\varphi}$ would have been sufficient. In Eq.\ (\ref{FunctionalwithoutD}) we have kept complex $\varphi$ and $\tilde{\varphi}$ in anticipation of the generalization to the case with diffraction. Now, using\ (\ref{FunctionalwithoutD}) with $F\lbrack\psi(\cdot)\rbrack =\delta(U-\vert\psi(L)\vert^2)$ in $p(U)=\left\langle\delta(U-\vert\psi(L)\vert^2)\right\rangle_S$, where $\langle\cdot\rangle_S$ denotes the average over the realizations of $S$, one gets
\begin{eqnarray}\label{actionintwithoutD}
&&p(U)=\int_{\varphi(0)=1}\left\langle\delta(U-\vert\varphi(L)\vert^2)
\, {\rm e}^{\frac{i}{2}\, \left(\left\langle\tilde{\varphi}\left\vert d_z -g\vert S\vert^2\right\vert\varphi\right\rangle
+c.\, c.\right)}\right\rangle_S \, \mathscr{D}^2\varphi\, \mathscr{D}^2\tilde{\varphi} \nonumber \\
&&=\int_{\varphi(0)=1}
\delta(U-\vert\varphi(L)\vert^2)\, {\rm e}^{\frac{i}{2}\, \left(\left\langle\tilde{\varphi}\left\vert d_z -g\vert S\vert^2\right\vert\varphi\right\rangle
+c.\, c.\right) -\left\langle S\left\vert T_C^{-1}\right\vert S\right\rangle}\, 
\mathscr{D}^2\varphi\, \mathscr{D}^2\tilde{\varphi}\, \mathscr{D}^2 S ,
\end{eqnarray}
where $T_C$ is the covariance operator of $S$ defined in Eq.\ (\ref{covariance1D}). The functional integral representation of $p(U)$ in Eq.\ (\ref{actionintwithoutD}) is of the same form as the one in Eq.\ (\ref{generalactionint}) with MSR action
\begin{eqnarray}\label{MSRactionwithoutD}
&&\mathcal{A}(\varphi,\tilde{\varphi},S)=
\frac{i}{2}\, \left(\left\langle\tilde{\varphi}\left\vert d_z -g\vert S\vert^2\right\vert\varphi\right\rangle
+c.\, c.\right) -\left\langle S\left\vert T_C^{-1}\right\vert S\right\rangle \nonumber \\
&&=\frac{i}{2}\, \left\lbrack\int_0^L \tilde{\varphi}^\ast(z) \left(d_z\varphi(z)
-g\vert S(z)\vert^2\varphi(z)\right) dz +c.\, c.\right\rbrack
-\int_0^L S^\ast(z) (T_C^{-1}S)(z)\, dz.
\end{eqnarray}
%
%
\subsection{Leading instanton and tail of $\bm{p(U)}$}\label{instantonwithoutD}
The leading instanton which determines the large $\ln U$ behavior of $p(U)$ in Eq.\ (\ref{actionintwithoutD}) is a stationary point of $\mathcal{A}(\varphi,\tilde{\varphi},S)$ under the restriction $\vert\varphi(L)\vert^2=U$. According to the usual procedure of Lagrange multipliers\ \cite{CH1989}, it can be found as a stationary point of the action $\mathcal{A}^{\prime}(\varphi,\tilde{\varphi},S)=\mathcal{A}(\varphi,\tilde{\varphi},S)+\lambda\vert\varphi(L)\vert^2$ without restriction, where $\lambda$ is a Lagrange multiplier. Write $\delta\mathcal{A}^{\prime}(\varphi,\tilde{\varphi},S)$ the variation of $\mathcal{A}^{\prime}(\varphi,\tilde{\varphi},S)$ under variations of the fields and their complex conjugates treated as independent variables, with endpoints $\varphi(0)=\varphi^\ast(0)=1$, and $\tilde{\varphi}(L^+)=\tilde{\varphi}^\ast(L^+)=0$ (by causality principle. See, e.g., Ref.\ \cite{FKLM1996}). It is convenient to make the independence of $\tilde{\varphi}$ and $\tilde{\varphi}^\ast$ explicit by writing $\tilde{\varphi}^\ast =\tilde{\vartheta}$, independent of $\tilde{\varphi}$. The stationarity condition $\delta\mathcal{A}^{\prime}(\varphi,\tilde{\varphi},S)=0$ leads to the equations
\begin{eqnarray}\label{insteqwithoutDphi1}
&&d_z\varphi(z)-g\vert S(z)\vert^2\varphi(z)=0\ {\rm with}\ \varphi(0)=1, \nonumber \\
&&d_z\tilde{\varphi}(z)+g\vert S(z)\vert^2\tilde{\varphi}(z)=-2i\lambda\varphi(L)\delta(z-L)
\ {\rm with}\ \tilde{\varphi}(L^+)=0, \\
&&d_z\tilde{\vartheta}(z) +g\vert S(z)\vert^2\tilde{\vartheta}(z) 
=-2i\lambda\varphi^\ast(L)\delta(z-L)
\ {\rm with}\ \tilde{\vartheta}(L^+)=0, \nonumber
\end{eqnarray}
and
\begin{equation}\label{insteqwithoutDS1}
(T_C^{-1}S)(z)=-\frac{ig}{2}\left\lbrack\tilde{\vartheta}(z)\varphi(z)
+\tilde{\varphi}(z)\varphi^\ast(z)\right\rbrack S(z),
\end{equation}
or, equivalently,
\begin{eqnarray}\label{insteqwithoutDphi2}
&&d_z\varphi(z)-g\vert S(z)\vert^2\varphi(z)=0\ {\rm with}\ \varphi(0)=1, \nonumber \\
&&d_z\tilde{\varphi}(z)+g\vert S(z)\vert^2\tilde{\varphi}(z)=0
\ {\rm with}\ \tilde{\varphi}(L)=2i\lambda\varphi(L),  \\
&&d_z\tilde{\vartheta}(z) +g\vert S(z)\vert^2\tilde{\vartheta}(z) =0
\ {\rm with}\ \tilde{\vartheta}(L) =2i\lambda\varphi^\ast(L) , \nonumber
\end{eqnarray}
and
\begin{equation}\label{insteqwithoutDS2}
\left\lbrack T_C\left(\tilde{\vartheta}\varphi +\tilde{\varphi}\varphi^\ast\right) S\right\rbrack (z)
=\frac{2i}{g}S(z).
\end{equation}
The equations\ (\ref{insteqwithoutDphi2}) are readily solved. One gets,
\begin{eqnarray}\label{instsolwithoutDphi1}
&&\varphi(z)={\rm e}^{g\int_0^z\vert S(z^\prime)\vert^2 dz^\prime} , \nonumber \\
&&\tilde{\varphi}(z)=2i\lambda\varphi(L){\rm e}^{g\int_z^L\vert S(z^\prime)\vert^2 dz^\prime} , \\
&&\tilde{\vartheta}(z)=2i\lambda\varphi^\ast(L){\rm e}^{g\int_z^L\vert S(z^\prime)\vert^2 dz^\prime} , \nonumber
\end{eqnarray}
and $\tilde{\vartheta}(z)\varphi(z)=\tilde{\varphi}(z)\varphi^\ast(z) =4i\lambda\vert\varphi(L)\vert^2=4i\lambda U$, independent of $z$. Injecting this solution onto the left-hand side of\ (\ref{insteqwithoutDS2}), one obtains the eigenvalue equation
\begin{equation}\label{insteqwithoutDS3}
(T_C S)(z)=\frac{1}{2\lambda gU} S(z).
\end{equation}
It follows immediately from Eq.\ (\ref{insteqwithoutDS3}) that an instanton solution for $S$ is an eigenfunction of its covariance operator $T_C$, which fixes the value of $\lambda$ for each instanton, namely $\lambda =1/(2\mu_n gU)$, where $\mu_1>\mu_2 >\cdots >0$ are the eigenvalues of $T_C$. Using\ (\ref{instsolwithoutDphi1}) and\ (\ref{insteqwithoutDS3}) in\ (\ref{MSRactionwithoutD}), one finds that the action of an instanton solution is $\mathcal{A} =-\mu_n^{-1}\| S\|_2^2$, and the leading instanton $S_{\rm inst}$ corresponds to the largest eigenvalue $\mu_1$ for which $\mathcal{A}$ is maximum. For simplicity, we assume that $\mu_1$ is not degenerate. The generalization to a degenerate $\mu_1$ is straightforward and we leave it to the reader as an exercice. Writing $\phi_1$ the normalized eigenfunction associated with $\mu_1$, one gets
\begin{equation}\label{instsolwithoutDS1}
S_{\rm inst}(z)= c_1 \phi_1(z),
\end{equation}
where $c_1$ is a complex number. The other components of the leading instanton, $\varphi_{\rm inst}$, $\tilde{\varphi}_{\rm inst}$, and $\tilde{\vartheta}_{\rm inst}$, are the instanton solution in Eq.\ (\ref{instsolwithoutDphi1}) with $S=S_{\rm inst}$ and $\lambda =1/(2\mu_1 gU)$. In the following, we will only need the expression of $\varphi_{\rm inst}(L)$, 
\begin{equation}\label{instsolwithoutDphi2}
\varphi_{\rm inst}(L)=\exp\left(g\| S_{\rm inst}\|_2^2\right) ,
\end{equation}
where $\|\cdot\|_2$ denotes the $L^2$-norm over $\lbrack 0,L\rbrack$.

Integrating out the fluctuations around the instanton, at fixed instanton, in Eq.~(\ref{actionintwithoutD}) and using the expressions of $S_{\rm inst}$ and $\varphi_{\rm inst}$ in Eqs.\ (\ref{instsolwithoutDS1}) and\ (\ref{instsolwithoutDphi2}), one obtains the diffraction-free version of the asymptotic expression\ (\ref{tailpofUgeneral}),
\begin{eqnarray}\label{tailpofUwithoutD1}
\ln p(U)&\sim&\ln\int\delta\left( U-{\rm e}^{2g\vert c_1\vert^2}\right)
\, {\rm e}^{-\vert c_1\vert^2/\mu_1}
\, \frac{d^2c_1}{\pi\mu_1} \nonumber \\
&=&\ln\int_0^{+\infty}\delta\left(U-{\rm e}^{2g\eta}\right)
\, {\rm e}^{-\eta/\mu_1}\frac{d\eta}{\mu_1}\ \ \ \ \ (\ln U\to +\infty),
\end{eqnarray}
where we have made the change of variable $\vert c_1\vert^2=\eta$. It can be seen in Eq.\ (\ref{tailpofUwithoutD1}) that $c_1$ is a (complex) Gaussian random variable with $\langle c_1\rangle =\langle c_1^2\rangle =0$ and $\langle\vert c_1\vert^2\rangle =\mu_1$. Thus, writing $c_1=\sqrt{\eta}\, {\rm e}^{i\arg(c_1)}$ in Eq.\ (\ref{instsolwithoutDS1}), one gets
\begin{equation}\label{instsolwithoutDS2}
S_{\rm inst}(z)= \sqrt{\eta}\, {\rm e}^{i\arg(c_1)} \phi_1(z),
\end{equation}
where $\eta$ is an exponential random variable with $p(\eta)=\mu_1^{-1}{\rm e}^{-\eta/\mu_1}$, and $\arg(c_1)$ is a random phase uniformly distributed over $\lbrack 0,2\pi)$. For large $\ln U$ (hence large $\eta$), the realizations of $S$ which contribute to the tail of $p(U)$ concentrate onto the leading instanton, $S(z)\sim S_{\rm inst}(z)$ ($\eta\to +\infty$), and one recovers the result of\ \cite{MD2004,MC2011} recalled in Eq.\ (\ref{oldresultwithoutD1}) (for $d_1=1$). It remains to perform the integral over $\eta$ in Eq.\ (\ref{tailpofUwithoutD1}), which can be done without difficulty. One obtains the asymptotic behavior given in Eq.\ (\ref{tailpofUwithoutD}) with $d_1=1$,
\begin{equation*}
\ln p(U)= -\left(1+\frac{1}{2\mu_1g}\right)\, \ln U +O(1)
\ \ \ \ \ (\ln U\to +\infty),
\end{equation*}
as expected.
%
%
\section{Amplification with diffraction: general setting}\label{withDsec}
The approach followed in the previous section to deal with the diffraction-free case is completely different from the one in\ \cite{MD2004,MC2011}. Having checked that both give the same results, we can now move on to the next step and use the instanton analysis to deal with the full problem with diffraction.
%
%
\subsection{MSR action and instanton equations}\label{actioninstantonwithD}
We consider the transversally one-dimensional ($d=1$) version of Eq.~(\ref{withDeq}),
\begin{equation}\label{withDeq1D}
\left\lbrace
\begin{array}{l}
\partial_z\psi(x,z)-\frac{i}{2m}\partial^2_{x^2} \psi(x,z)=g\vert S(x,z)\vert^2\psi(x,z), \\
0\le z\le L,\ x\in\Lambda\subset\mathbb{R},\ {\rm and}\ \psi(x,0)=1,
\end{array}\right.
\end{equation}
where we take for $\Lambda$ the circle of length $\ell$. The random field $S(x,z)$ is homogeneous along $x$ with normalization $L^{-1}\int_0^L \langle\vert S(x,z)\vert^2\rangle\, dz =1$. (The generalization to more than one transverse dimension is straightforward.) Our goal is to determine the realizations of $S(x,z)$ and the tail of $p(U)$ in the large $\ln U$ limit for $U=\vert\psi(0,L)\vert^2$, with $\psi(x,z)$ solution to Eq.~(\ref{withDeq1D}). Write
\begin{equation}\label{paraxialoperator}
D_{z,\, x^2}\equiv\partial_z -\frac{i}{2m}\partial^2_{x^2},
\end{equation}
and $T_C$ the covariance operator of $S$ defined by
\begin{equation}\label{covariance3D1}
(T_C f)(x,z)=\int_\Lambda\int_0^L
C(x-x^\prime ,z,z^\prime)\, f(x^\prime ,z^\prime)\, dz^\prime\, dx^\prime ,
\ \ \ f(x,z)\in L^2(\Lambda\times [0,L]),
\end{equation}
with $C(x-x^\prime ,z,z^\prime)=\langle S(x,z)S(x^\prime ,z^\prime)^\ast \rangle$. The counterpart of Eq.\ (\ref{actionintwithoutD}) in the problem with diffraction reads
\begin{equation}\label{actionintwithD}
p(U)=\int_{\varphi(x,0)=1}
\delta\left(U-\vert\varphi(0,L)\vert^2\right)\, {\rm e}^{\frac{i}{2}\, \left(\left\langle\tilde{\varphi}\left\vert D_{z,\, x^2}-g\vert S\vert^2\right\vert\varphi\right\rangle
+c.\, c.\right) -\left\langle S\left\vert T_C^{-1}\right\vert S\right\rangle}\, 
\mathscr{D}^2\varphi\, \mathscr{D}^2\tilde{\varphi}\, \mathscr{D}^2 S,
\end{equation}
which is of the same form as in Eq.\ (\ref{generalactionint}) with MSR action
\begin{eqnarray}\label{MSRactionwithD}
&&\mathcal{A}(\varphi,\tilde{\varphi},S)=
\frac{i}{2}\, \left(\left\langle\tilde{\varphi}\left\vert D_{z,\, x^2} -g\vert S\vert^2\right\vert\varphi\right\rangle
+c.\, c.\right) -\left\langle S\left\vert T_C^{-1}\right\vert S\right\rangle \nonumber \\
&&=\frac{i}{2}\, \left\lbrack\int_{\Lambda}\int_0^L \tilde{\varphi}^\ast(x,z) \left(D_{z,\, x^2}\varphi(x,z)
-g\vert S(x,z)\vert^2\varphi(x,z)\right) dz\, dx +c.\, c.\right\rbrack \\
&&-\int_{\Lambda}\int_0^L S^\ast(x,z) (T_C^{-1}S)(x,z)\, dz\, dx. \nonumber
\end{eqnarray}
The derivation of the instanton equations from the action in Eq.\ (\ref{MSRactionwithD}) follows exactly the same line as in the diffraction-free case in Sec.\ \ref{instantonwithoutD}. Varying $\mathcal{A}(\varphi,\tilde{\varphi},S)$ with the Lagrange multiplier term $\lambda\vert\varphi(0,L)\vert^2$ and setting the variation to zero, one obtains the equations
\begin{eqnarray}\label{insteqwithDphi1}
&&\left\lbrack D_{z,\, x^2}-g\vert S(x,z)\vert^2\right\rbrack\varphi(x,z)=0
\ {\rm with}\ \varphi(x,0)=1, \nonumber \\
&&\left\lbrack D_{z,\, x^2}+g\vert S(x,z)\vert^2\right\rbrack\tilde{\varphi}(x,z)=-2i\lambda\varphi(0,L)\delta(x)\delta(z-L)
\ {\rm with}\ \tilde{\varphi}(x,L^+)=0, \\
&&\left\lbrack D_{z,\, x^2}^\ast +g\vert S(x,z)\vert^2\right\rbrack\tilde{\vartheta}(x,z)=-2i\lambda\varphi^\ast(0,L) \delta(x)\delta(z-L)
\ {\rm with}\ \tilde{\vartheta}(x,L^+)=0, \nonumber
\end{eqnarray}
and
\begin{equation}\label{insteqwithDS1}
(T_C^{-1}S)(x,z)=-\frac{ig}{2}\left\lbrack\tilde{\vartheta}(x,z)\varphi(x,z)
+\tilde{\varphi}(x,z)\varphi^\ast(x,z)\right\rbrack S(x,z),
\end{equation}
or, equivalently,
\begin{eqnarray}\label{insteqwithDphi2}
&&\left\lbrack D_{z,\, x^2}-g\vert S(x,z)\vert^2\right\rbrack\varphi(x,z)=0
\ {\rm with}\ \varphi(x,0)=1, \nonumber \\
&&\left\lbrack D_{z,\, x^2}+g\vert S(x,z)\vert^2\right\rbrack\tilde{\varphi}(x,z)=0
\ {\rm with}\ \tilde{\varphi}(x,L)=2i\lambda\varphi(0,L)\delta(x),  \\
&&\left\lbrack D_{z,\, x^2}^\ast +g\vert S(x,z)\vert^2\right\rbrack\tilde{\vartheta}(x,z)=0
\ {\rm with}\ \tilde{\vartheta}(x,L) =2i\lambda\varphi^\ast(0,L) \delta(x) , \nonumber
\end{eqnarray}
and
\begin{equation}\label{insteqwithDS2}
\left\lbrack T_C\left(\tilde{\vartheta}\varphi +\tilde{\varphi}\varphi^\ast\right) S\right\rbrack (x,z)
=\frac{2i}{g}S(x,z).
\end{equation}
The equations\ (\ref{insteqwithDphi2}) are readily solved in terms of Feynman-Kac propagator,
\begin{equation}\label{FKpropagator}
K(x_2,z_2;x_1,z_1)=\int_{x(z_1)=x_1}^{x(z_2)=x_2}{\rm e}^{\int_{z_1}^{z_2}
\left\lbrack\frac{im}{2}\dot{x}(\tau)^2+g\vert S(x(\tau),\tau)\vert^2\right\rbrack\, d\tau}
\mathscr{D}x,
\end{equation}
with $z_2>z_1$, where the path-integral is over the set of all the continuous paths in $\Lambda$ satisfying $x(z_1)=x_1$ and $x(z_2)=x_2$. One gets
\begin{eqnarray}\label{instsolwithDphi1}
&&\varphi(x,z)=\int_\Lambda K(x,z;y,0)\, dy, \nonumber \\
&&\tilde{\varphi}(x,z)=2i\lambda\varphi(0,L)\, K(0,L;x,z)^\ast , \\
&&\tilde{\vartheta}(x,z)=2i\lambda\varphi^\ast(0,L)\, K(0,L;x,z) . \nonumber
\end{eqnarray}
Using the expressions\ (\ref{instsolwithDphi1}) on the left-hand side of\ (\ref{insteqwithDS2}), one obtains
\begin{equation}\label{insteqwithDS3}
\varphi^\ast(0,L)\, G_1(x,z)+\varphi(0,L)\, G_2(x,z)=\frac{1}{\lambda g}\, S(x,z),
\end{equation}
with
\begin{eqnarray}\label{integralG1}
G_1(x,z)=&&\int_0^L\int_\Lambda\int_\Lambda
K(0,L;x^\prime,z^\prime)K(x^\prime,z^\prime;\xi,0) \nonumber \\
&&\times C(x-x^\prime,z,z^\prime)\, S(x^\prime,z^\prime)
\, dx^\prime\, d\xi\, dz^\prime ,
\end{eqnarray}
and
\begin{eqnarray}\label{integralG2}
G_2(x,z)=&&\int_0^L\int_\Lambda\int_\Lambda
K(0,L;x^\prime,z^\prime)^\ast K(x^\prime,z^\prime;\xi,0)^\ast \nonumber \\
&&\times C(x-x^\prime,z,z^\prime)\, S(x^\prime,z^\prime)
\, dx^\prime\, d\xi\, dz^\prime .
\end{eqnarray}

In the large $\ln U$ limit, $S$ concentrates onto the leading instanton, $S_{\rm inst}$, solution to Eq.~(\ref{insteqwithDS3}), and $\varphi(x,z)$ concentrates onto $\varphi_{\rm inst}(x,z)$ given by the Feynman-Kac path-integral for $\varphi(x,z)$ in Eq.~(\ref{instsolwithDphi1}) with $S=S_{\rm inst}$. For a given $C$ it is always possible, in principle, to solve the instanton equation~(\ref{insteqwithDS3}) numerically. To this end, an iterative forward-backward scheme as introduced, e.g., in\ \cite{CS2001} could be used to solve the equivalent system (\ref{insteqwithDphi2})-(\ref{insteqwithDS2}). However, the fact that $\ln U$ is large allows Eq.~(\ref{insteqwithDS3}) to be solved analytically (in this limit) without the need to specify $C$. The key is to call upon the well-known gain narrowing effect\ \cite{YL1963}, here in the space of continuous paths $x(\cdot)$, according to which the Feynman-Kac propagators in\ (\ref{integralG1})-(\ref{integralG2}) are dominated by the contribution of the paths with the largest amplification, the contribution of the other paths being subdominant in the large amplification limit. These dominant trajectories run in the vicinity of `ridge paths' along which $\vert S_{\rm inst}(x,z)\vert^2$ is at a global maximum for every given $z$. Assuming that the ridge paths are all continuous (to be checked {\it a posteriori}, once $S_{\rm inst}$ is known), Eq.~(\ref{insteqwithDS3}) simplifies and for a wide class of $S$ that we will now specify, it can be solved explicitly.
%
%
\subsection{Specification of $\bm{S(x,z)}$}\label{modelanddef}
We assume that $S(x,z)$ can be expressed as a finite random Fourier sum,
\begin{equation}\label{KLexpansion}
S(x,z)=\sum_{(n,j)\in\mathcal{I}}s_{(n,j)}\sqrt{\frac{\sigma_{(n,j)}}{\ell}}\, 
{\rm e}^{2i\pi nx/\ell}\Phi_{(n,j)}(z),
\end{equation}
where $\mathcal{I}$ is a finite subset of $\mathbb{Z}\times\mathbb{N}$. The $s_{(n,j)}$s are complex Gaussian random variables with $\langle s_{(n,j)}\rangle =\langle s_{(n,j)}s_{(m,k)}\rangle =0$ and $\langle s_{(n,j)}s_{(m,k)}^\ast\rangle =\delta_{nm}\delta_{jk}$, the $\sigma_{(n,j)}$s are positive constants normalized to $\sum_{(n,j)\in\mathcal{I}}\sigma_{(n,j)}=L\ell$, and, for fixed $n$, the $\Phi_{(n,j)}$s are orthonormal continuous functions of $0\le z\le L$. Using Eq.\ (\ref{KLexpansion}) in $C(x-x^\prime ,z,z^\prime)=\langle S(x,z)S(x^\prime ,z^\prime)^\ast \rangle$ one gets
\begin{equation}\label{CFexpansion}
C(x-x^\prime ,z,z^\prime)=
\sum_{(n,j)\in\mathcal{I}}\frac{\sigma_{(n,j)}}{\ell}\, {\rm e}^{2i\pi n(x-x^\prime)/\ell}
\Phi_{(n,j)}(z)\Phi_{(n,j)}(z^\prime)^\ast ,
\end{equation}
from which it follows that $\sigma_{(n,j)}$ and ${\rm e}^{2i\pi nx/\ell}\Phi_{(n,j)}(z)/\sqrt{\ell}$ are the eigenvalues and orthonormal eigenfunctions of the covariance operator $T_C$ defined in Eq.\ (\ref{covariance3D1}).

Equation\ (\ref{KLexpansion}) generalizes models of spatially smoothed laser beams in which laser light is represented by a superposition of monochromatic beamlets the amplitudes of which are independent random variables\ \cite{RD1993}. For a large number of beamlets these random variables can be taken as Gaussian and the laser electric field takes on the form\ (\ref{KLexpansion}) in which the sum over $j$ reduces to $j=1$ with $\Phi_{(n,1)}(z)=(1/\sqrt{L})\, \exp\lbrack i\alpha (2\pi n/\ell)^2z)\rbrack$ where $\alpha$ is a (real) constant. Moreover, every centered Gaussian field with a continuous correlation function has an expansion of the form\ (\ref{KLexpansion}), possibly with an infinite sum\ \cite{AT2007}. Combining this result with the practically unavoidable existence of some natural cut-off making the sum finite (like, e.g., in numerical simulations), one can safely expects the expression in Eq.\ (\ref{KLexpansion}) to be quite generic, at least from a practical point of view.

Let $B(0,L)$ denote the set of all the continuous paths in $\Lambda$ satisfying $x(L)=0$ and define $M\lbrack x(\cdot)\rbrack$ the $\vert\mathcal{I}\vert\times\vert\mathcal{I}\vert$ positive definite matrix with components
\begin{equation}\label{matrixM}
M_{(n,j)(m,k)}\lbrack x(\cdot)\rbrack =\frac{\sqrt{\sigma_{(n,j)}\sigma_{(m,k)}}}{\ell}
\int_0^L{\rm e}^{2i\pi(m-n)x(z)/\ell}\Phi_{(n,j)}(z)^\ast\Phi_{(m,k)}(z)\, dz,
\end{equation}
in which $x(\cdot)\in B(0,L)$. Since $M\lbrack x(\cdot)\rbrack$ is positive definite, all its eigenvalues are real and positive. Write $\mu_1\lbrack x(\cdot)\rbrack >0$ the largest eigenvalue of $M\lbrack x(\cdot)\rbrack$. It is proved in\ \cite{MCL2006} that the eigenvalues of $M\lbrack x(\cdot)\rbrack$ are equal to the ones of $T_{x(\cdot)}$ defined by
\begin{equation}\label{covariance3Dpath}
(T_{x(\cdot)} f)(z) =\int_0^L
C(x(z)-x(z^\prime) ,z,z^\prime)\, f(z^\prime)\, dz^\prime .
\ \ \ f(z)\in L^2([0,L]).
\end{equation}
It follows in particular that $\mu_1\lbrack x(\cdot)\rbrack$ is invariant under the path transformations leaving $C(x(z)-x(z^\prime),z,z^\prime)$ unchanged and that the image of a path maximizing $\mu_1\lbrack x(\cdot)\rbrack$ by such a transformation is also a path maximizing $\mu_1\lbrack x(\cdot)\rbrack$. We consider cases fulfilling the following two assumptions:
\bigskip

(i) all the paths maximizing $\mu_1\lbrack x(\cdot)\rbrack$ are in $B(0,L)$;
\bigskip

(ii) there is a finite number of paths in $B(0,L)$ maximizing $\mu_1\lbrack x(\cdot)\rbrack$.
\bigskip

\noindent
Assumption (i) is a central feature of the class of $S$ we consider in this paper. We don't know whether Eq.~(\ref{insteqwithDS3}) could be solved analytically in the large $\ln U$ limit without this assumption. The technical restriction~(ii) will be used in Sec.~\ref{singlepathtail}. Lifting (ii) --- e.g., in the case of an uncountable set of maximizing paths --- raises tricky technical problems yet to be solved; this will be the subject of a future work. Both assumptions (i) and (ii) are fulfilled in most cases of practical interest. Finally, for notational convenience we define
\begin{equation}\label{mumax}
\mu_{\rm max}=\sup_{x(\cdot)\in B(0,L)}\mu_1\lbrack x(\cdot)\rbrack ,
\end{equation}
the supremum being reached in $B(0,L)$, by Assumption (i).
%
%
\section{Single-filament instanton and tail of $\bm{p(U)}$}\label{singlepathsec}
In the following, we consider the simplest case where for each realization of $S_{\rm inst}$ there is only one ridge path of $\vert S_{\rm inst}(x,z)\vert^2$ in $B(0,L)$, denoted by $x_{\rm inst}(\cdot)$, dominating the large $\ln U$ limit of the Feynman-Kac propagators in (\ref{integralG1}) and (\ref{integralG2}). Note that $x_{\rm inst}(\cdot)$ may be different from one realization of $S_{\rm inst}$ to the other. The set of all the realizations of $S_{\rm inst}$ having the same $x_{\rm inst}(\cdot)$ defines a random field, denoted by $S_{\rm inst}^{x_{\rm inst}(\cdot)}$, referred to in the following as a `single-filament instanton' (the reason for this name will appear more clearly at the end of Sec.~\ref{singlepathLI}). We will write $\varphi_{\rm inst}^{x_{\rm inst}(\cdot)}(x,z)$ the Feynman-Kac path-integral for $\varphi(x,z)$ in Eq.~(\ref{instsolwithDphi1}) with $S=S_{\rm inst}^{x_{\rm inst}(\cdot)}$.

Single-filament instantons are not the only possible solutions to Eq.~(\ref{insteqwithDS3}). Multi-filament instantons are also possible if realizations of $\vert S_{\rm inst}(x,z)\vert^2$ have more than one ridge path. The conditions for single- or multi-filament instantons are specified below Eq.~(\ref{instsolwithDS1}) as well as at the end of Appendix\ \ref{app1}. The study of multi-filament instantons being excessively intricate, we restrict ourselves to single-filament instantons for the sake of clarity and readability.
%
%
\subsection{Leading instanton}\label{singlepathLI}
Assume that $x_{\rm inst}(\cdot)$ is continuous (to be checked {\it a posteriori}). As finite sums of continuous functions, both $S(x,z)$ and $C(x-x^\prime ,z,z^\prime)$ are continuous functions of their arguments. It follows in particular that for fixed $x$, $z$, and $z^\prime$, the product $C(x-x^\prime,z,z^\prime)\, S(x^\prime,z^\prime)$ on the right-hand side of Eqs.~(\ref{integralG1}) and (\ref{integralG2}) is a continuous function of $x^\prime$. Integrating over $\xi$ and $x^\prime$ at fixed $z^\prime$ and using the fact that, in the large $\ln U$ limit, only the vicinity of $x^\prime =x_{\rm inst}(z^\prime)$ contributes, one gets the large $\ln U$ behavior of $G_{1,2}(x,z)$,
\begin{eqnarray}\label{integralG1largeU}
&&G_1(x,z)\sim\int_\Lambda K(0,L;\xi,0)\, d\xi
\int_0^L C(x-x_{\rm inst}(z^\prime),z,z^\prime)
\, S_{\rm inst}^{x_{\rm inst}(\cdot)}(x_{\rm inst}(z^\prime),z^\prime)\, dz^\prime \\
&&=\varphi_{\rm inst}^{x_{\rm inst}(\cdot)}(0,L)
\, \int_0^L C(x-x_{\rm inst}(z^\prime),z,z^\prime)
\, S_{\rm inst}^{x_{\rm inst}(\cdot)}(x_{\rm inst}(z^\prime),z^\prime)\, dz^\prime
\ \ \ \ \ (\ln U\to +\infty), \nonumber
\end{eqnarray}
and
\begin{eqnarray}\label{integralG2largeU}
&&G_2(x,z)\sim\int_\Lambda K(0,L;\xi,0)^\ast\, d\xi
\int_0^L C(x-x_{\rm inst}(z^\prime),z,z^\prime)
\, S_{\rm inst}^{x_{\rm inst}(\cdot)}(x_{\rm inst}(z^\prime),z^\prime)\, dz^\prime \\
&&=\varphi(0,L)_{\rm inst}^{x_{\rm inst}(\cdot)\, \ast}
\, \int_0^L C(x-x_{\rm inst}(z^\prime),z,z^\prime)
\, S_{\rm inst}^{x_{\rm inst}(\cdot)}(x_{\rm inst}(z^\prime),z^\prime)\, dz^\prime
\ \ \ \ \ (\ln U\to +\infty). \nonumber
\end{eqnarray}
Injecting these expressions onto the left-hand side of Eq.\ (\ref{insteqwithDS3}), one obtains the instanton equation
\begin{equation}\label{insteqwithDS3largeU1}
\int_0^L C(x-x_{\rm inst}(z^\prime),z,z^\prime)
\, S_{\rm inst}^{x_{\rm inst}(\cdot)}(x_{\rm inst}(z^\prime),z^\prime)\, dz^\prime
=\frac{1}{2\lambda gU}\, S_{\rm inst}^{x_{\rm inst}(\cdot)}(x,z),
\end{equation}
where we have used the equality $\vert\varphi_{\rm inst}^{x_{\rm inst}(\cdot)}(0,L)\vert^2=U$ imposed by the delta function on the right-hand side of (\ref{actionintwithD}).

Equation\ (\ref{insteqwithDS3largeU1}) can be solved in two different ways, depending on wether or not the Fourier decompositions (\ref{KLexpansion}) and (\ref{CFexpansion}) for $S_{\rm inst}^{x_{\rm inst}(\cdot)}$ and $C(x-x_{\rm inst}(z^\prime),z,z^\prime)$ are used in Eq.~(\ref{insteqwithDS3largeU1}). Using these decompositions, one gets the eigenvalue equation
\begin{equation}\label{insteqwithDS3largeU2}
\sum_{(m,k)\in\mathcal{I}}M_{(n,j)(m,k)}\lbrack x_{\rm inst}(\cdot)\rbrack\, s_{(m,k)}=
\frac{1}{2\lambda gU}\, s_{(n,j)},
\end{equation}
which fixes $\lambda$ at $\lambda =1/(2\mu_n\lbrack x_{\rm inst}(\cdot)\rbrack gU)$, where $\mu_1\lbrack x_{\rm inst}(\cdot)\rbrack >\mu_2\lbrack x_{\rm inst}(\cdot)\rbrack >\cdots >0$ are the eigenvalues of $M\lbrack x_{\rm inst}(\cdot)\rbrack$. Using (\ref{instsolwithDphi1}) and (\ref{insteqwithDS3largeU1}) (or (\ref{insteqwithDS3largeU2})) in (\ref{MSRactionwithD}), one finds that the action of an instanton solution is $\mathcal{A} =-\mu_n\lbrack x_{\rm inst}(\cdot)\rbrack^{-1}\| S(x_{\rm inst}(\cdot),\cdot)\|_2^2$, and the leading instanton $S_{\rm inst}^{x_{\rm inst}(\cdot)}$ for which $\mathcal{A}$ is maximum corresponds to the largest eigenvalue $\mu_1\lbrack x_{\rm inst}(\cdot)\rbrack$ with $x_{\rm inst}(\cdot)$ maximizing $\mu_1\lbrack x(\cdot)\rbrack$. Note that $x_{\rm inst}(\cdot)$ is a non-random path. The fact that $x_{\rm inst}(\cdot)$ exists and is continuous is ensured by the assumption (i). Thus, for every path $x_{\rm inst}(\cdot)\in B(0,L)$ maximizing $\mu_1[x(\cdot)]$, there is a leading instanton
\begin{equation}\label{instsolwithDS1}
S_{\rm inst}^{x_{\rm inst}(\cdot)}(x,z)=\sum_{(n,j)\in\mathcal{I}}
\mathfrak{s}_{(n,j)}\sqrt{\frac{\sigma_{(n,j)}}{\ell}}\, 
{\rm e}^{2i\pi nx/\ell}\Phi_{(n,j)}(z),
\end{equation}
where $\bm{\mathfrak{s}}$ (with components $\mathfrak{s}_{(n,j)}$) is an eigenvector of $M\lbrack x_{\rm inst}(\cdot)\rbrack$ associated with the largest eigenvalue $\mu_1\lbrack x_{\rm inst}(\cdot)\rbrack$. It is checked in Appendix\ \ref{app1} that $x_{\rm inst}(\cdot)$ is indeed a ridge path of $\vert S_{\rm inst}^{x_{\rm inst}(\cdot)}(x,z)\vert^2$, as it should be.

The calculation in Appendix\ \ref{app1} also specifies under what condition a leading instanton is a single-filament instanton. Namely, the fundamental eigenspace of $M\lbrack x_{\rm inst}(\cdot)\rbrack$ and the one of  $M\lbrack x(\cdot)\rbrack$ for every other path maximizing $\mu_1\lbrack x(\cdot)\rbrack$, if any, must be essentially disjoint\footnote{`essentially disjoint' and `trivial intersection' mean that the intersection reduces to the zero vector.}. In particular, if the fundamental eigenspaces of $M\lbrack x(\cdot)\rbrack$ for all the different paths maximizing $\mu_1\lbrack x(\cdot)\rbrack$ are essentially disjoint, all the instantons are single-filament instantons. This is the case considered in this paper. Conversely, if the fundamental eigenspaces of $M\lbrack x(\cdot)\rbrack$ for different paths maximizing $\mu_1\lbrack x(\cdot)\rbrack$ have a non trivial intersection, then multi-filament instantons come into play as possible solutions to Eq.~(\ref{insteqwithDS3}).

We now solve the equation\ (\ref{insteqwithDS3largeU1}) without using the Fourier decompositions (\ref{KLexpansion}) and~(\ref{CFexpansion}). It can be seen from Eq.~(\ref{insteqwithDS3largeU1}) that $S_{\rm inst}^{x_{\rm inst}(\cdot)}(x_{\rm inst}(z),z)$ is an eigenfunction of $T_{x_{\rm inst}(\cdot)}$ with eigenvalue $1/2\lambda gU$, hence $\lambda =1/(2\mu_n\lbrack x_{\rm inst}(\cdot)\rbrack gU)$, where $\mu_1\lbrack x_{\rm inst}(\cdot)\rbrack >\mu_2\lbrack x_{\rm inst}(\cdot)\rbrack >\cdots >0$ are the eigenvalues of $T_{x_{\rm inst}(\cdot)}$. (Recall that $T_{x(\cdot)}$ and $M\lbrack x(\cdot)\rbrack$ have the same eigenvalues with the same multiplicities \cite{MCL2006}). As explained below Eq.~(\ref{insteqwithDS3largeU2}), $S_{\rm inst}^{x_{\rm inst}(\cdot)}$ corresponds to the largest eigenvalue $\mu_1\lbrack x_{\rm inst}(\cdot)\rbrack$ with $x_{\rm inst}(\cdot)$ maximizing $\mu_1\lbrack x(\cdot)\rbrack$. Like in Sec.~\ref{instantonwithoutD}, we assume for simplicity that $\mu_{\rm max}=\mu_1\lbrack x_{\rm inst}(\cdot)\rbrack$ is not degenerate. The case of a degenerate $\mu_{\rm max}$ can be dealt with similarly without difficulties (the calculation is more technical and the results not substantially different). Writing $\phi_1$ the normalized fundamental eigenmode of $T_{x_{\rm inst}(\cdot)}$, one has
\begin{equation}\label{instsolwithDSalongpath}
S_{\rm inst}^{x_{\rm inst}(\cdot)}(x_{\rm inst}(z),z)= c_1 \phi_1(z),
\end{equation}
where $c_1$ is a complex number. Injecting (\ref{instsolwithDSalongpath}) into (\ref{insteqwithDS3largeU1}), one obtains
\begin{equation}\label{instsolwithDS2}
S_{\rm inst}^{x_{\rm inst}(\cdot)}(x,z)=\frac{c_1}{\mu_{\rm max}}
\, \int_0^L C(x-x_{\rm inst}(z^\prime),z,z^\prime)
\, \phi_1(z^\prime)\, dz^\prime .
\end{equation}
%

\bigskip
\noindent$\bullet$ {\it Equivalence of the Fourier and convolution representations of $S_{\rm inst}^{x_{\rm inst}(\cdot)}$}

The fact that the expressions of $S_{\rm inst}^{x_{\rm inst}(\cdot)}(x,z)$ in Eqs.~(\ref{instsolwithDS1}) and (\ref{instsolwithDS2}) are equivalent is proved in Appendix\ \ref{app2}, with $\mathfrak{s}_{(n,j)}$ in Eq.~(\ref{instsolwithDS1}) and $c_1$ in Eq.~(\ref{instsolwithDS2}) related to each other by
\begin{equation}\label{stoc-ctos}
\mathfrak{s}_{(n,j)}=\frac{1}{\sqrt{\mu_{\rm max}}}
\, c_1 \mathfrak{e}^{(1)}_{(n,j)}
\ {\rm and}\ 
c_1=\sqrt{\mu_{\rm max}}
\, \sum_{(n,j)\in\mathcal{I}}\mathfrak{s}_{(n,j)}\mathfrak{e}^{(1)\, \ast}_{(n,j)},
\end{equation}
where $\bm{\mathfrak{e}}^{(1)}$ is the normalized fundamental eigenvector of $M\lbrack x_{\rm inst}(\cdot)\rbrack$ defined by its components
\begin{equation}\label{vector-e}
\mathfrak{e}^{(1)}_{(n,j)}=
\sqrt{\frac{\sigma_{(n,j)}}{\ell\, \mu_{\rm max}}}\, 
\int_0^L{\rm e}^{-2i\pi nx_{\rm inst}(z^\prime)/\ell}\Phi_{(n,j)}(z^\prime)^\ast
\phi_1(z^\prime)\, dz^\prime .
\end{equation}
Note that by Eq.~(\ref{instsolwithDS1}) for $S_{\rm inst}^{x_{\rm inst}(\cdot)}(x_{\rm inst}(z),z)$ and the definition of $\mathfrak{e}^{(1)}_{(n,j)}$ in Eq.~(\ref{vector-e}), the expression of $c_1$ in Eq.~(\ref{stoc-ctos}) coincides with $c_1 =\int_0^L S_{\rm inst}^{x_{\rm inst}(\cdot)}(x_{\rm inst}(z),z)\phi_1(z)^\ast dz$ in agreement with Eq.~(\ref{instsolwithDSalongpath}), as it should be.
%

\bigskip
\noindent$\bullet$ {\it Statistical properties of $S$ in the large $\ln U$ limit}

The statistical properties of $c_1$ and the $\mathfrak{s}_{(n,j)}$s are readily obtained from the ones of the $s_{(n,j)}$s in Eq.~(\ref{KLexpansion}). Since the $s_{(n,j)}$s are i.i.d. standard complex Gaussian random variables with $\langle s_{(n,j)}\rangle =\langle s_{(n,j)}^2\rangle =0$ and $\langle\vert s_{(n,j)}\vert^2\rangle =1$, the orthogonal projection of the $\vert\mathcal{I}\vert$-dimensional (complex) vector $\bm{s}$ with coordinates $s_{(n,j)}$ onto any given direction is also a standard complex Gaussian random variable statistically independent of the projections onto the orthogonal directions. By (\ref{stoc-ctos}), $c_1$ is proportional to the projection of $\bm{s}$ onto the direction of the vector $\bm{\mathfrak{e}}^{(1)}$ defined in (\ref{vector-e}), from which it follows that $c_1$ is a complex Gaussian random variable with $\langle c_1\rangle =\langle c_1^2\rangle =0$ and $\langle\vert c_1\vert^2\rangle =\mu_{\rm max}$. The statistical properties of the $\mathfrak{s}_{(n,j)}$s are different from the ones of the $s_{(n,j)}$s because $\bm{\mathfrak{s}}$ is restricted to the one-dimensional fundamental eigenspace of $M\lbrack x_{\rm inst}(\cdot)\rbrack$, which induces correlations between the $\mathfrak{s}_{(n,j)}$s. From the first Eq.~(\ref{stoc-ctos}) and the statistical properties of $c_1$, one finds that the $\mathfrak{s}_{(n,j)}$s are correlated complex Gaussian random variables with $\langle \mathfrak{s}_{(n,j)}\rangle =\langle \mathfrak{s}_{(n,j)}\mathfrak{s}_{(m,k)}\rangle =0$ and $\langle \mathfrak{s}_{(n,j)}\mathfrak{s}_{(m,k)}^\ast\rangle =\mathfrak{e}^{(1)}_{(n,j)}\mathfrak{e}^{(1)\, \ast}_{(m,k)}$.

Normalizing $S$ to its $L^2$-norm, Equation~(\ref{instsolwithDS2}) and $S\sim S_{\rm inst}^{x_{\rm inst}(\cdot)}$ ($\ln U\to +\infty$) yield
\begin{equation}\label{quasideterministicprofile}
\frac{S(x,z)}{\| S\|_2}\sim A{\rm e}^{i\arg(c_1)} \int_0^L C(x-x_{\rm inst}(z^\prime),z,z^\prime)
\, \phi_1(z^\prime)\, dz^\prime\ \ \ \ \ (\ln U\to +\infty),
\end{equation}
where $\|\cdot\|_2$ denotes the $L^2$-norm over $\Lambda\times\lbrack 0,L\rbrack$, $A$ is a positive constant, and $\arg(c_1)$ is a random phase uniformly distributed over $\lbrack 0,2\pi)$. From Eq.~(\ref{quasideterministicprofile}) it follows immediately that $\vert S(x,z)\vert/\| S\|_2$ is non-random, which means that the profile of $S(x,z)$ is purely deterministic in this case. This result generalizes the diffraction-free deterministic profile of $S$ in Eq.~(\ref{oldresultwithoutD2}) when diffraction is taken into account.
%

\bigskip
\noindent$\bullet$ {\it Typical shape of $S_{\rm inst}^{x_{\rm inst}(\cdot)}$}

Although the Fourier representation of $S_{\rm inst}^{x_{\rm inst}(\cdot)}$ in Eq.~(\ref{instsolwithDS1}) is very useful to deal with technical points like in Appendix\ \ref{app1}, it is not so clear as to the structure of $S_{\rm inst}^{x_{\rm inst}(\cdot)}$ in real space. By contrast, it is easier to figure out the shape of $S_{\rm inst}^{x_{\rm inst}(\cdot)}(x,z)$ from the convolution representation in Eq.~(\ref{instsolwithDS2}), knowing the correlation function $C(x-x^\prime ,z,z^\prime)$. As a simple illustration, take, e.g., $C(x-x^\prime ,z,z^\prime)=f[(x-x^\prime)/x_c ,(z-z^\prime)/z_c]$, where $x_c$ and $z_c$ respectively denote transverse and axial correlation lengths, $f(x,z)$ being negligibly small outside the domain defined by both $\vert x\vert\le 1$ and $\vert z\vert\le 1$. Assuming $x_c\ll\ell$, $z_c\ll L$ and a `gentle' ridge path with $\vert\dot{x}_{\rm inst}(z)\vert\lesssim\ell/L$ for all $0\le z\le L$, it is not difficult to show from Eq.~(\ref{instsolwithDS2}) that $S_{\rm inst}^{x_{\rm inst}(\cdot)}$ lives within a thin tube, or filament, of radius $\rho\lesssim x_c+\ell z_c/L\ll\ell$ along the path $x_{\rm inst}(\cdot)$. This is the reason for the name `single-filament instanton' given to $S_{\rm inst}^{x_{\rm inst}(\cdot)}$. An example of the elongated profile of $S_{\rm inst}^{x_{\rm inst}(\cdot)}$ can be seen in Fig.~\ref{figureinst} (Sec.~\ref{numerics}).
%
%
\subsection{Tail of $\bm{p(U)}$}\label{singlepathtail}
Write $N_{\rm inst}$ the number of single-filament instantons. $N_{\rm inst}$ is the number of paths maximizing $\mu_1\lbrack x(\cdot)\rbrack$, which is finite by assumption (ii). As mentioned below (\ref{instsolwithDS1}), we consider cases where the fundamental eigenspaces of $M\lbrack x(\cdot)\rbrack$ for all the different paths maximizing $\mu_1\lbrack x(\cdot)\rbrack$ are essentially disjoint. It means that the instantons --- that are all single-filament instantons --- are mutually exclusive realizations of $S$. As a result, the total instanton contribution to the tail of $p(U)$ in Eq.~(\ref{actionintwithD}) is the sum of the $N_{\rm inst}$ individual instanton contributions. Let $\pi^{(i)}(U)$ denotes the contribution of the $i$th instanton. It will be seen below that the leading term of $\ln\pi^{(i)}(U)$ in the large $U$ limit does not depend on $i$. Writing $\ln f(U)$ this term and $\pi^{(i)}(U)=f(U)A^{(i)}(U)$ with $\ln A^{(i)}(U)=o[\ln f(U)]$ as $\ln U\to +\infty$, one has
\begin{eqnarray}\label{tailpofUwithD1}
\ln p(U)&\sim&\ln\sum_{i=1}^{N_{\rm inst}}\pi^{(i)}(U)
=\ln\sum_{i=1}^{N_{\rm inst}}f(U)A^{(i)}(U) \nonumber \\
&=&\ln f(U)+\ln\sum_{i=1}^{N_{\rm inst}}A^{(i)}(U) \\
&=&\ln f(U)+o[\ln f(U)]\ \ \ \ \ (\ln U\to +\infty). \nonumber
\end{eqnarray}
Thus, at leading order, $\ln p(U)\sim\ln f(U)$ where $\ln f(U)$ is the leading term of $\ln\pi^{(i)}(U)$ for all $1\le i\le N_{\rm inst}$. Let $\lbrack {\rm LT}\rbrack^{(U\to +\infty)}(\cdot)$ denote the leading term of the asymptotic expansion of $(\cdot)$ as $\ln U\to +\infty$. Picking a $i$ and integrating out the fluctuations of the fields around the corresponding $i$th single-filament instanton (with ridge-path $x_{\rm inst}(\cdot)$) in (\ref{actionintwithD}), one obtains
\begin{eqnarray}\label{tailpofUwithD2}
&&\ln f(U)=\lbrack {\rm LT}\rbrack^{(U\to +\infty)}
\ln\left\langle\delta(U-\vert\varphi_{\rm inst}^{x_{\rm inst}(\cdot)}(0,L)\vert^2)\right\rangle_{S_{\rm inst}^{x_{\rm inst}(\cdot)}} \\
&&=\lbrack {\rm LT}\rbrack^{(U\to +\infty)}
\ln\, \left\lbrack\int\delta\left( U-\vert\varphi_{\rm inst}^{x_{\rm inst}(\cdot)}(0,L)\vert^2\right)
\, {\rm e}^{-\vert c_1\vert^2/\mu_{\rm max}}\, \frac{d^2c_1}{\pi\mu_{\rm max}}
\right\rbrack . \nonumber
\end{eqnarray}
To go further we need the behavior of $\vert\varphi_{\rm inst}^{x_{\rm inst}(\cdot)}(0,L)\vert^2$ as a function of $c_1$ in the large $\ln U$ limit. First, we make the change of variables $c_1 =\sqrt{\eta}\, {\rm e}^{i\arg(c_1)}$, where $\eta$ is an exponential random variable with $p(\eta)=\mu_{\rm max}^{-1}{\rm e}^{-\eta/\mu_{\rm max}}$, and $\arg(c_1)$ is a random phase uniformly distributed over $\lbrack 0,2\pi)$. For finite $L$ and $\ell$, a large $\ln U$ implies a large $\eta$. It is clear from Eq.~(\ref{instsolwithDS2}) and the first Eq.~(\ref{instsolwithDphi1}) with $S=S_{\rm inst}^{x_{\rm inst}(\cdot)}$ that $\vert\varphi_{\rm inst}^{x_{\rm inst}(\cdot)}(0,L)\vert^2$ does not depend on $\arg(c_1)$. Writing
\begin{equation}\label{expressionforphi}
\vert\varphi_{\rm inst}^{x_{\rm inst}(\cdot)}(0,L)\vert^2=
A(\eta)\, {\rm e}^{2g\eta},
\end{equation}
without loss of generality, on the right-hand side of Eq.~(\ref{tailpofUwithD2}), one obtains
\begin{eqnarray}\label{tailpofUwithD3}
&&\ln f(U)=\lbrack {\rm LT}\rbrack^{(U\to +\infty)}
\ln\int_0^{+\infty}
\delta\left(U-A(\eta)\, {\rm e}^{2g\eta}\right)
\, p(\eta)\, d\eta \nonumber \\
&&=\lbrack {\rm LT}\rbrack^{(U\to +\infty)}
\ln\int_0^{+\infty}
\frac{\delta\left(\eta-\eta(U)\right)}{\vert 2g+\partial_{\eta}\ln A(\eta)\vert\, U}
\, p(\eta)\, d\eta \\
&&=\lbrack {\rm LT}\rbrack^{(U\to +\infty)}
\ln\frac{1}{U^{1+1/2\mu_{\rm max}g}}
\frac{\eta(U)\, A(\eta(U))^{1/2\mu_{\rm max}g}}
{\vert 2g+\partial_{\eta}\ln A(\eta(U))\vert} \nonumber
\end{eqnarray}
where $\eta(U)$ is the solution to $A(\eta)\, {\rm e}^{2g\eta}=U$. It is shown in Appendix\ \ref{app3} that
\begin{equation}\label{limitsofA}
\begin{array}{l}
\lim_{\eta\to +\infty}\frac{1}{\eta}\, \ln A(\eta)=0, \\
\lim_{\eta\to +\infty}\partial_{\eta}\ln A(\eta)=0,
\end{array}
\end{equation}
from which it follows that $\eta(U)\sim (2g)^{-1}\ln U$ and $\vert 2g+\partial_{\eta}\ln A(\eta(U))\vert\sim 2g$ in Eq.~(\ref{tailpofUwithD3}) which reads
\begin{eqnarray}\label{tailpofUwithD4}
\ln f(U)&=&\lbrack {\rm LT}\rbrack^{(U\to +\infty)}
\ln\left(\frac{A((2g)^{-1}\ln U)^{1/2\mu_{\rm max}g}
\, \ln U}{4g^2\, U^{1+1/2\mu_{\rm max}g}}
\right) \nonumber \\
&=& -\left(1+\frac{1}{2\mu_{\rm max}g}\right)\, \ln U,
\end{eqnarray}
where we have used $\ln A((2g)^{-1}\ln U)^{1/2\mu_{\rm max}g}=o(\ln U)$ (see the end of appendix \ref{app3}). Note that the expression of $\ln f(U)$ in Eq.~(\ref{tailpofUwithD4}) is independent of $i$, as announced. Using (\ref{tailpofUwithD4}) on the right-hand side of (\ref{tailpofUwithD1}), one finally obtains the tail of $p(U)$ as
\begin{equation}\label{tailpofUwithD5}
\ln p(U)=-\left(1+\frac{1}{2\mu_{\rm max}g}\right)\, \ln U+o(\ln U)\ \ \ \ \ (\ln U\to +\infty),
\end{equation}
from which it follows that $p(U)$ has a leading algebraic tail $\propto U^{-\zeta}$ modulated by a slow varying amplitude (slower than algebraic) with exponent $\zeta =(1+1/2\mu_{\rm max} g)$. Injecting this result into $\langle U\rangle =\int_{1}^{+\infty}Up(U)\, dU$, one finds that the critical coupling in the case of single-filament instantons is given by $g_c(L)=1/2\mu_{\rm max}$, where $\mu_{\rm max}$ depends on $L$, in agreement with the general result of Ref.\ \cite{MCL2006} recalled in Eq.~(\ref{critcouplingwithD}).
%
%
\section{Numerical results}\label{numerics}
In this section we present the results of numerical simulations for realizations of $S$ in a sample of experimentally realistic size. Of particular interest are laser-plasma interaction experiments with space-time optical smoothing in which $S$ is renewed periodically. For typical parameters on current large laser facilities, the number of uncorrelated realizations of $S$ generated after $1$ to $10$  shots of a $10$~ns laser beam with coherence time between $0.5$~ps and $1$~ps is between $10^4$ and $2\ 10^5$, which sets the size of the sample we consider here. Since the number of realizations is limited, whether or not asymptotic results can be sampled is essential to get an idea of what to expect --- and not to expect --- from the simulations. As a Gaussian field, $S$ has a fast decreasing probability at large $\vert S\vert$ which makes $p(\ln U\gg 1)$ so small that $\ln U\gg 1$ is out of reach of the sample we consider. So we don't expect to directly check the asymptotic instanton in (\ref{instsolwithDS2}) match the numerical results, or the tail of $p(U)$ in (\ref{tailpofUwithD5}) match the numerical histogram of $U$. However, asymptotic behavior comes out gradually as $U$ increases and instantons beginning to emerge from the noise may already affect the results in the accessible, not asymptotic regime. If so, simulations should show it and, hopefully, provide useful information on the transition to the asymptotic regime.

One possible way to reach the asymptotic regime would be to bias the underlying distribution of $S$ towards the outcomes of interest, like, e.g., in the `importance sampling algorithm'\ \cite{HM1956} frequently used in rare event physics (see e.g.\ \cite{HMS2019} and references therein). In principle, this approach should allow the tail of $p(U)$ to be probed at high $\ln U\gg 1$. A direct numerical test of our analytical results, which is important from a theoretical viewpoint but of a somewhat lesser interest experimentally, will be the subject of a future work.

For the sake of completeness, note that the comparison of asymptotic instanton solutions to results of numerical simulations has been carried out for nonlinear equations with additive noise in several works. See, e.g.,\ \cite{GGS2013,HMS2019} and references therein for the Burgers and Kardar-Parisi-Zhang  equations, respectively.

For definiteness, we have taken
\begin{equation}\label{solSnum}
S(x,z)=\sum_{n=-50}^{50}s_{n}\sqrt{\varsigma_n}\, 
\exp\, i\left\lbrack\frac{2\pi n}{\ell} x+\left(\frac{2\pi n}{\ell}\right)^2\frac{z}{2}\right\rbrack ,
\end{equation}
where the $s_n$s are complex Gaussian random variables with $\langle s_n\rangle =\langle s_n s_m\rangle =0$ and $\langle s_n s_m^\ast\rangle =\delta_{nm}$, and the spectral density $\varsigma_n$ is normalized to $\sum_{n=-50}^{50}\varsigma_n=1$. Equation~(\ref{solSnum}) --- which is of the form\ (\ref{KLexpansion}) in which the sum over $j$ reduces to $j=1$ and $\varsigma_n =\sigma_{(n,1)}/L\ell$ --- is reminiscent of models of spatially smoothed laser beams\ \cite{RD1993}, where $S$ is a solution to the paraxial wave equation
\begin{equation}\label{eqSnum}
\partial_z S(x,z)+\frac{i}{2}\partial^2_{x^2} S(x,z)=0,
\end{equation}
here with boundary condition $S(x,0)=\sum_{n=-50}^{50} s_n\sqrt{\varsigma_n}\, \exp(2i\pi nx/\ell)$. To ensure that the space average $\ell^{-1}\int_{\Lambda}S(x,z)\, dx$ is zero for all $z$ and every realization of $S$, as expected for the electric field of a smoothed laser beam, the mode at $n=0$ is excluded by taking $\varsigma_{0}=0$. Here we show the results for the Gaussian spectrum
\begin{equation}\label{gaussspectrnum}
\varsigma_{n\ne 0}\propto\exp\left\lbrack -\left(\frac{\pi n}{\ell}\right)^2\right\rbrack .
\end{equation}
(Other widely used spectra, like top-hat and Cauchy spectra, give similar results.)

For each realization of $S$ on a cylinder of length $L=10$ and circumference $\ell =20$, we have solved Eq.~(\ref{withDeq}) by using a symmetrized $z$-split method \cite{Strang1968} which propagates the diffraction term, $(i/2m)\partial^2_{x^2}\psi(x,z)$, in Fourier space and the amplification term, $g\vert S(x,z)\vert^2\psi(x,z)$, in real space. We have taken $m=0.7$, and $g=0.5$. To get a better statistics of large amplification values, we have considered $U_{\rm max}=\vert\psi(x_{\rm max},L)\vert^2$ instead of $U=\vert\psi(0,L)\vert^2$, where $x_{\rm max}$ is the value of $x$ maximizing $\vert\psi(x,L)\vert^2$ (i.e., the location of the highest peak of $\vert\psi(x,L)\vert^2$). As $\ln U_{\rm max}$ increases, $S$ is expected to concentrate onto the leading instanton(s) arriving at $x(L)\simeq x_{\rm max}$.

\begin{figure}[!h]
\includegraphics[width = 0.9\linewidth]{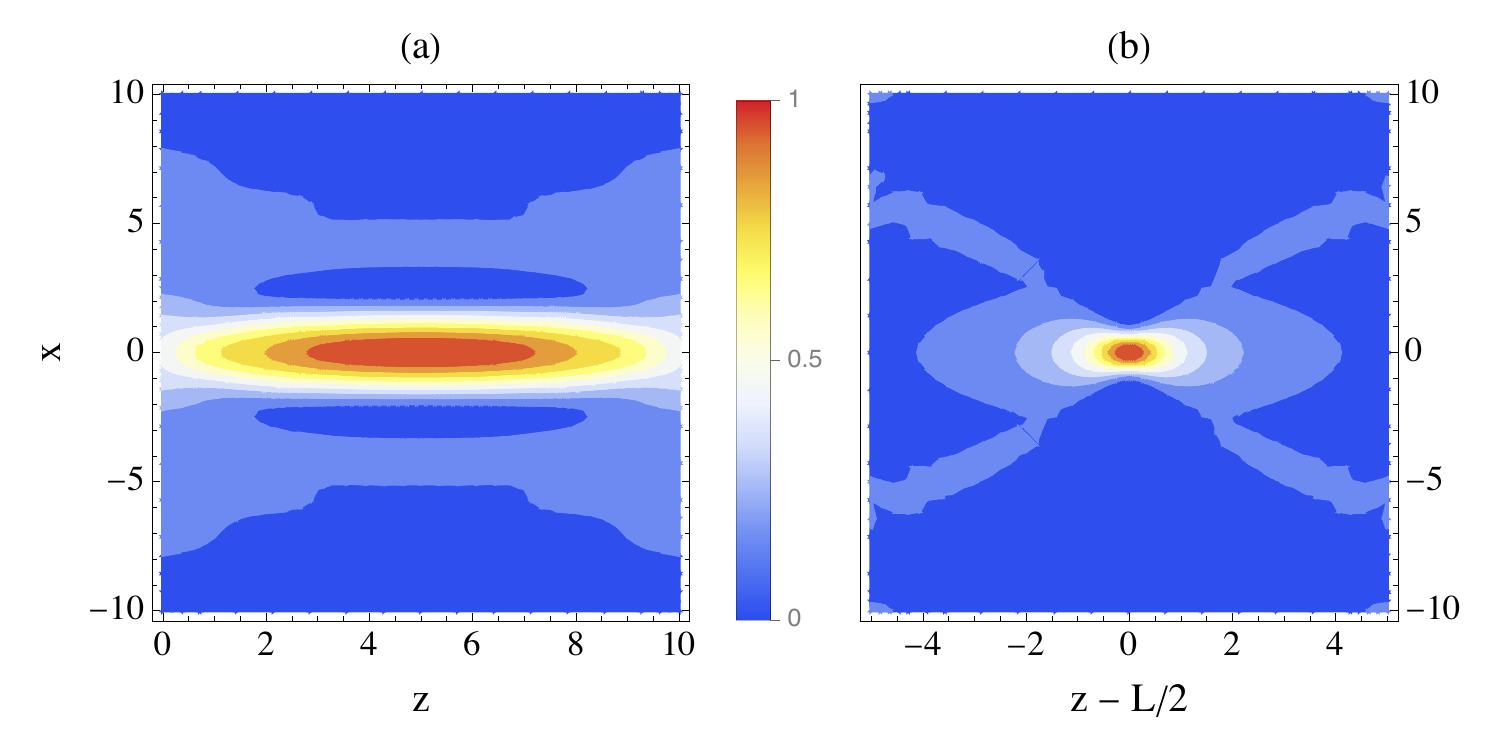}
\caption{(a) Contour plot of $\vert S_{\rm inst}(x,z)\vert^2$ normalized to $\max_{\Lambda\times\lbrack 0,L\rbrack}\vert S_{\rm inst}(x,z)\vert^2=1$. (b) Contour plot of the `hot spot profile' $\vert C(x,z)\vert^2$.}\label{figureinst}
\end{figure}

Write $S_{\rm inst}(x,z)$ the leading instanton(s) arriving at $x(L)=0$. For a given $x(\cdot)$ arriving at $x(L)=0$ we can compute the elements of $M\lbrack x(\cdot)\rbrack$, then its largest eigenvalue $\mu_1\lbrack x(\cdot)\rbrack$, numerically. Maximizing the result over fractional polynomial paths numerically, we found a unique global maximum at $x_{\rm inst}(\cdot)\equiv 0$ with non-degenerate $\mu_{\rm max}=4.34984$. We have then determined $S_{\rm inst}$ from Eqs.~(\ref{instsolwithDS1}), (\ref{solSnum}), and (\ref{gaussspectrnum}), with numerically computed eigenvector $\bm{\mathfrak{s}}$. The critical coupling is $g_c(L)=1/(2\mu_{\rm max})\simeq~0.11495$ and $g=0.5$ is in the above critical regime with $g/g_c(L)\simeq 4.35>1$. Figure~\ref{figureinst} shows the contour plots of $\vert S_{\rm inst}\vert^2$ and the `hot spot profile' $\vert C\vert^2$\ \cite{RD1993,G1999}.

\begin{figure}[!h]
\includegraphics[width = 0.6\linewidth]{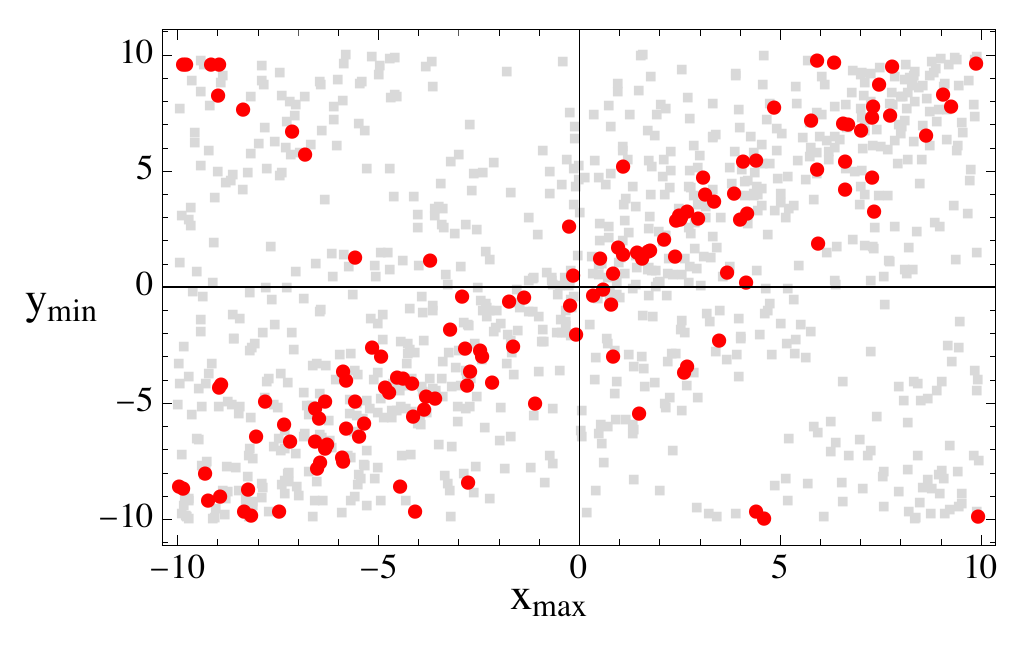}
\caption{Scatter plot of $x_{\rm max}$ and $y_{\rm min}$ for the realizations in $\lbrace S\rbrace$ with $\log_{10}U_{\rm max}\ge 18$. Red circles and gray squares are realizations in $\lbrace S\rbrace_{1\%}$ ($\mathfrak{D}\le 0.7$) and $\lbrace S\rbrace\setminus\lbrace S\rbrace_{1\%}$ ($\mathfrak{D}> 0.7$), respectively.}\label{figurecor}
\end{figure}

By statistical invariance under $x$-translation, the leading instanton arriving at $x(L)=y$ is $S_{\rm inst}^{y}(x,z)=S_{\rm inst}(x-y,z)$. Define $\hat{S}_{\rm inst}^{y}= S_{\rm inst}^{y}/\| S_{\rm inst}^{y}\|_{2,\Lambda\times\lbrack 0,L\rbrack}$ and $\hat{S}=S/\| S\|_{2,\Lambda\times\lbrack 0,L\rbrack}$ where $\|\cdot\|_{2,\Lambda\times\lbrack 0,L\rbrack}$ is the $L^2$-norm on $\Lambda\times\lbrack 0,L\rbrack$. Write $S_{\parallel}^{y}=\left(\hat{S}_{\rm inst}^{y},\hat{S}\right) \hat{S}_{\rm inst}^{y}$ the component of $\hat{S}$ along $S_{\rm inst}^{y}$. We have measured the difference between $S$ and the instanton through the minimized $L^2$-distance
\begin{eqnarray}\label{distnum1}
\mathfrak{D}\equiv d_2\left(\hat{S},S_{\parallel}^{y_{\rm min}}\right)&=& \min_{y\in\Lambda}\, 
\|\hat{S}-S_{\parallel}^{y}\|_{2,\Lambda\times\lbrack 0,L\rbrack} \nonumber \\
&=&\sqrt{1-\max_{y\in\Lambda}\, \vert(\hat{S},\hat{S}_{\rm inst}^{y})\vert^2},
\end{eqnarray}
where $y_{\rm min}$ is the value of $y$ minimizing $\|\hat{S}-S_{\parallel}^{y}\|_{2,\Lambda\times\lbrack 0,L\rbrack}$. The smaller $\mathfrak{D}$, the closer $S$ to the instanton arriving at $x(L)=y_{\rm min}$. The fact that $y_{\rm min}$ can be different from $x_{\rm max}$ is due to the fluctuations of $S$ away from the instanton, the relative amplitude of which is measured by $\mathfrak{D}$. For $\mathfrak{D}$ smaller than average, $y_{\rm min}\simeq x_{\rm max}$ with a relatively small dispersion of the data points about $y_{\rm min}=x_{\rm max}$, as can be seen in Fig.~\ref{figurecor}. Using the Fourier representations (\ref{solSnum}) for both $S$ and $S^y_{\rm Inst}$ on the right-hand side of Eq.~(\ref{distnum1}), one gets
\begin{equation}\label{distnum2}
\mathfrak{D}=\sqrt{1-
\frac{\max_{y\in\Lambda}\, \vert\sum_{n}\varsigma_n \hat{s}_n\hat{\mathfrak{s}}_n^\ast
{\rm e}^{2i\pi ny/\ell}\vert^2}{\left(\sum_{n}\varsigma_n \vert s_n\vert^2\right)\, 
\left(\sum_{n}\varsigma_n \vert \mathfrak{s}_n\vert^2\right)}},
\end{equation}
which is the expression we have used in the simulations.
\begin{figure}[!h]
\includegraphics[width = 0.6\linewidth]{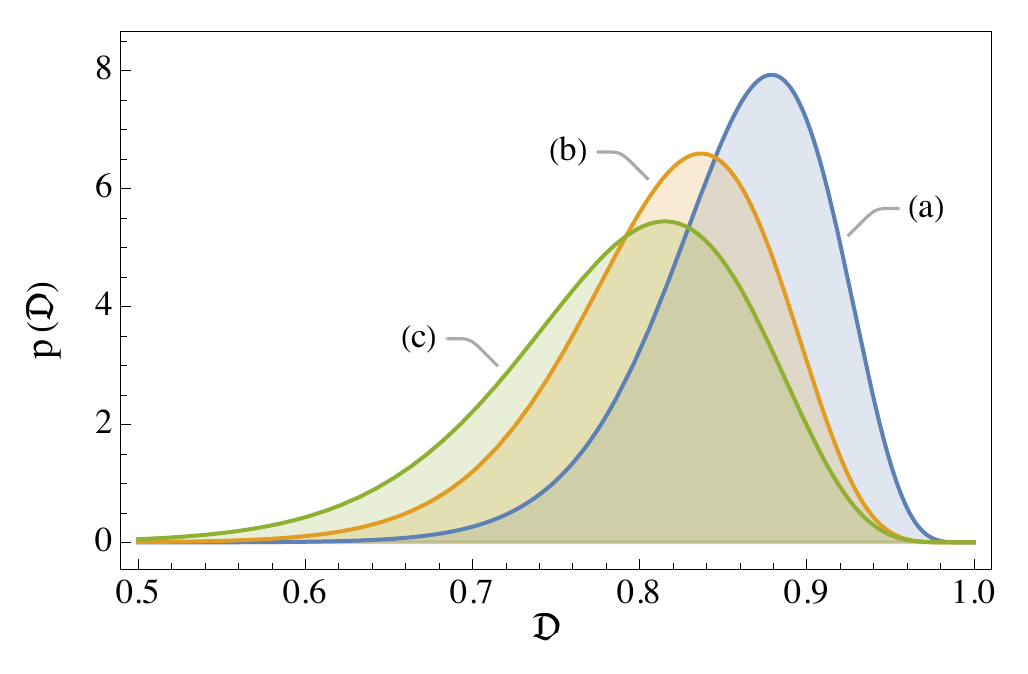}
\caption{Probability distribution of $\mathfrak{D}$ estimated from (a) $\lbrace S\rbrace$, (b) and (c) the realizations in $\lbrace S\rbrace$ with $\ln U_{\rm max}$ above the $90$th and $99$th percentiles, respectively.}\label{figureprobad}
\end{figure}
\begin{figure}[!h]
\includegraphics[width = 0.6\linewidth]{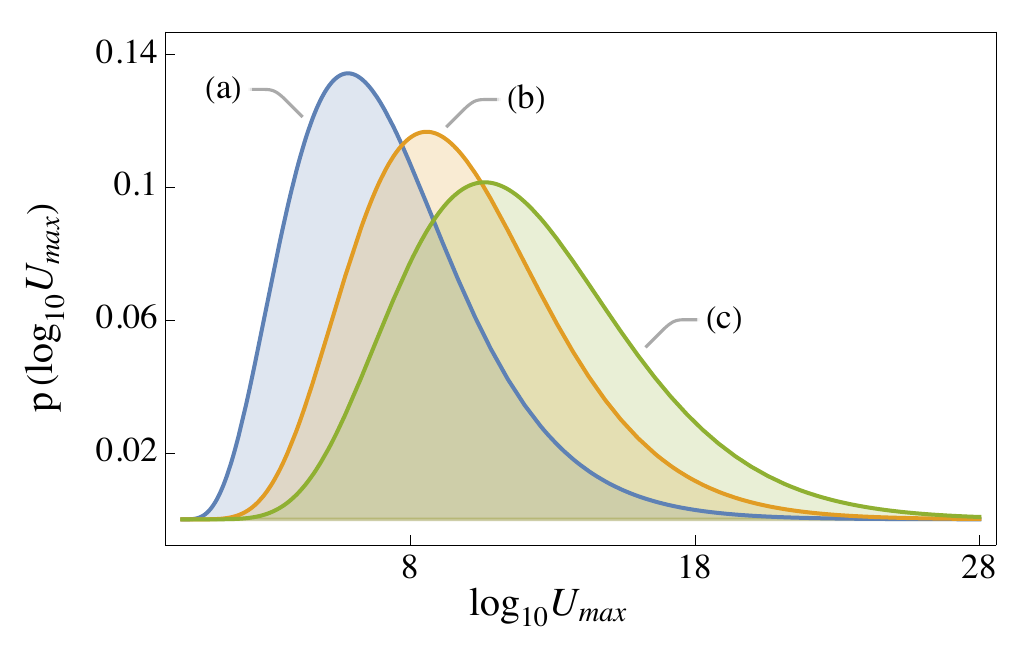}
\caption{Probability distribution of $\log_{10}U_{\rm max}$ estimated from (a) $\lbrace S\rbrace$, (b) $\lbrace S\rbrace_{10\%}$, and (c) $\lbrace S\rbrace_{1\%}$.}\label{figureprobau}
\end{figure}

We drew $10^5$ independent realizations of $S$ denoted in the following by $\lbrace S\rbrace$. Figure~\ref{figureprobad} shows the probability distribution of $\mathfrak{D}$ estimated from $\lbrace S\rbrace$ and the realizations in $\lbrace S\rbrace$ with $\ln U_{\rm max}$ above the $90$th and $99$th percentiles. The last two are conditional probabilities knowing that $\log_{10}U_{\rm max}\ge 12.7$ and $\log_{10}U_{\rm max}\ge 18$, respectively. One can see a clear tendency of $\mathfrak{D}$ to decrease with increasing $\log_{10}U_{\rm max}$: the subsamples of $\lbrace S\rbrace$ conditioned on a larger $\log_{10}U_{\rm max}$ are statistically biased toward the instanton compared with the unconditioned sample $\lbrace S\rbrace$ itself. This numerical result is consistent with the predicted concentration of $S$ onto the instanton for $\ln U\to +\infty$.

To study this bias in more detail, we have used the two samples $\lbrace S\rbrace_{10\%}$ and $\lbrace S\rbrace_{1\%}$ respectively defined as the realizations in $\lbrace S\rbrace$ with $\mathfrak{D}$ below the $10$th and $1$st percentiles. These samples correspond to $\mathfrak{D}\le 0.8$ for $\lbrace S\rbrace_{10\%}$ and $\mathfrak{D}\le 0.7$ for $\lbrace S\rbrace_{1\%}$. Figure~\ref{figureprobau} shows the probability distribution of $\log_{10}U_{\rm max}$ estimated from (a) $\lbrace S\rbrace$, (b) $\lbrace S\rbrace_{10\%}$, and (c) $\lbrace S\rbrace_{1\%}$. The last two are conditional probabilities knowing that $\mathfrak{D}\le 0.8$ and $\mathfrak{D}\le 0.7$, respectively. In this figure, the statistical bias of $S$, already observed in Fig.~\ref{figureprobad}, appears as the clear tendency of $\log_{10}U_{\rm max}$ to increase with decreasing $\mathfrak{D}$.
\begin{figure}[ht]
\includegraphics[width = 0.6\linewidth]{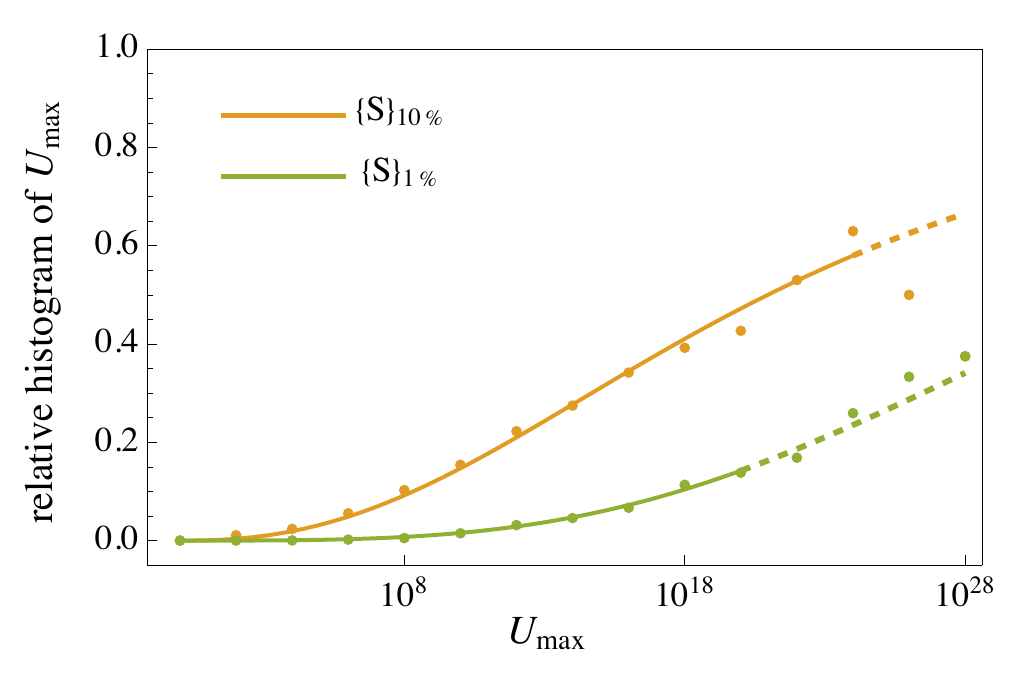}
\caption{Percentage of realizations of $S$ in $\lbrace S\rbrace_{10\%}$ (orange, upper curve) and $\lbrace S\rbrace_{1\%}$ (green, lower curve) for $\log_{10}U_{\rm max}$ in $\lbrack n ,\, n+2)$, with $n$ varying from $0$ to $28$. As guides to the eyes, the solid lines are nonlinear fits of the corresponding data points. Dashed lines are continuations of these fits to higher $\log_{10}U_{\rm max}$ (disregarding the data points in this domain)}\label{figureratio}
\end{figure}

The concentration of $S$ onto the instanton implies that for all $\varepsilon >0$ and $0<a<b$, one has $\lim_{a\to +\infty}{\rm Prob.}(\mathfrak{D}\le\varepsilon \vert\, a\le\log_{10}U_{\rm max}<b)=1$. Thus, for $a$ large enough it is not unreasonable to expect ${\rm Prob.}(\mathfrak{D}\le\varepsilon \vert\, a\le\log_{10}U_{\rm max}<b)$ to increase with increasing $a$, which should be possible to check numerically. ${\rm Prob.}(\mathfrak{D}\le\varepsilon \vert\, a\le\log_{10}U_{\rm max}<b)$ can be estimated by the percentage of realizations of $S$ with $\mathfrak{D}\le\varepsilon$ among the realizations with $a\le\log_{10}U_{\rm max}<b$. In Figure~\ref{figureratio} we show the results for $\varepsilon =0.8$ and $0.7$ (i.e., $S$ in $\lbrace S\rbrace_{10\%}$ and $\lbrace S\rbrace_{1\%}$, respectively), $a=n$, and $b=n+2$, with $n$ an integer. It can be seen that both curves increase with increasing $\log_{10}U_{\rm max}$, as expected. Note, e.g., that $30\%$ of the realizations with $\log_{10}U_{\rm max}\simeq 28$ are in $\lbrace S\rbrace_{1\%}$ (i.e., have $\mathfrak{D}\le 0.7$), when $\lbrace S\rbrace_{1\%}$ represents only $1\%$ of all the realizations in $\lbrace S\rbrace$: the emergence of a statistical bias of $S$ with increasing amplification is clearly visible.

Figure~\ref{figurecor} shows a scatter plot of $x_{\rm max}$ and $y_{\rm min}$ for the $10^3$ realizations in $\lbrace S\rbrace$ with the largest $U_{\rm max}$ (viz., $\log_{10} U_{\rm max}\ge 18$). Red circles and gray squares correspond to realizations with $\mathfrak{D}\le 0.7$ and $\mathfrak{D}> 0.7$, respectively (i.e., realizations in $\lbrace S\rbrace_{1\%}$ and $\lbrace S\rbrace\setminus\lbrace S\rbrace_{1\%}$). It can be checked that the dispersion of the data points about $y_{\rm min}=x_{\rm max}$ is indeed smaller for smaller $\mathfrak{D}$, as announced below Eq.~(\ref{distnum1}). Note that due to the periodic boundary condition in $\Lambda$, the distance between $x_{\rm max}$ and $y_{\rm min}$ is $\min(\vert x_{\rm max}-y_{\rm min}\vert ,\ell -\vert x_{\rm max}-y_{\rm min}\vert)$ and the data points in the left-upper and right-lower corners are actually close to the diagonal.
\begin{figure}[!h]
\includegraphics[width = 0.9\linewidth]{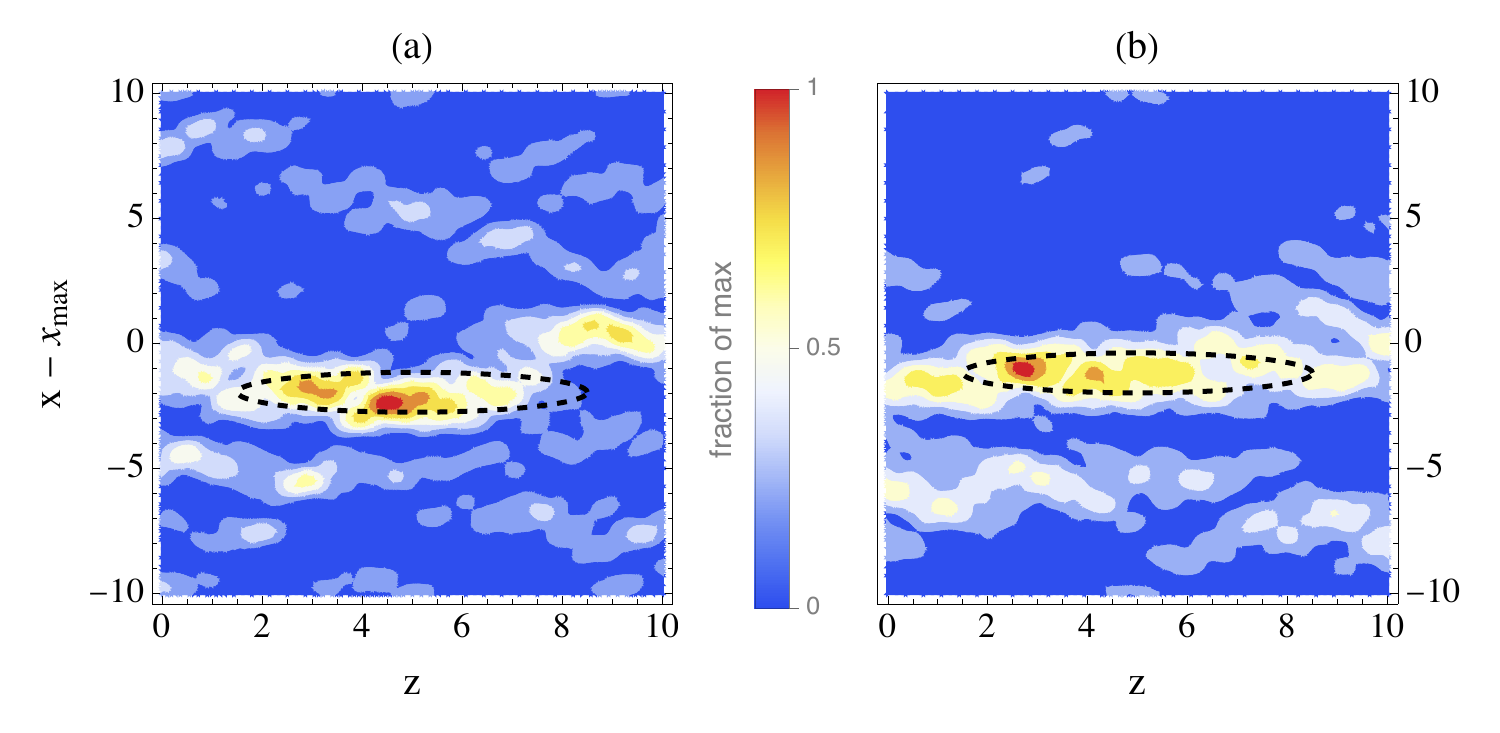}
\caption{Contour plots of $\vert S(x,z)\vert^2$ for two realizations in $\lbrace S\rbrace_{1\%}$ ($\mathfrak{D}\le 0.7$) with $\log_{10}U_{\rm max}\ge 27$. The dashed contours indicate the theoretical instanton arriving at $x(L)=y_{\rm min}$ for the considered realization. (a): $\log_{10}U_{\rm max}=29.63$, $\mathfrak{D}=0.68$, and $\max_{\Lambda\times\lbrack 0,L\rbrack}\vert S(x,z)\vert^2=14.83$. (b): $\log_{10}U_{\rm max}=27$, $\mathfrak{D}=0.537$, and $\max_{\Lambda\times\lbrack 0,L\rbrack}\vert S(x,z)\vert^2=13$.}\label{figuresmalld}
\end{figure}
\begin{figure}[!h]
\includegraphics[width = 0.9\linewidth]{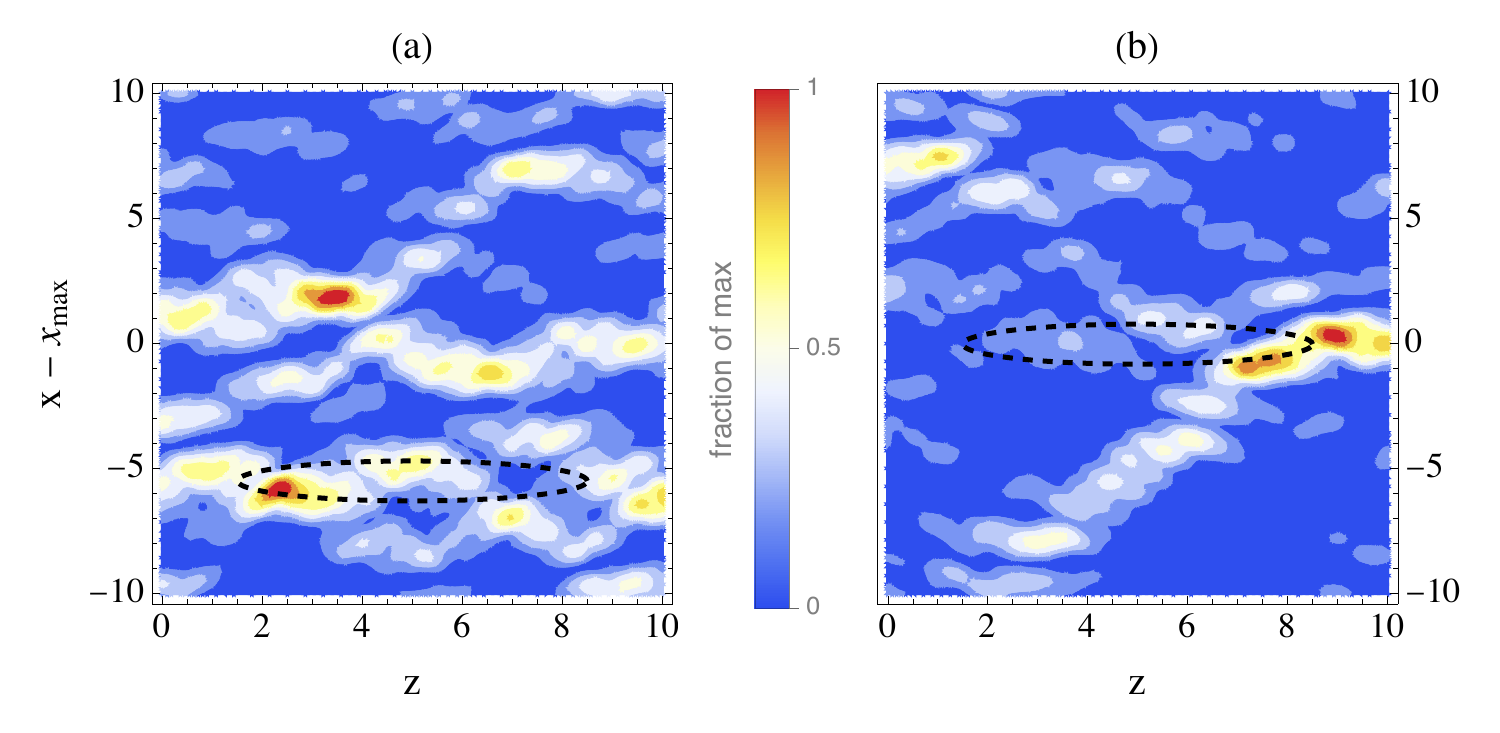}
\caption{Plots similar to the ones shown in Figure \ref{figuresmalld} for two realizations in $\lbrace S\rbrace\setminus\lbrace S\rbrace_{1\%}$ ($\mathfrak{D}> 0.7$). (a): $\log_{10}U_{\rm max}=29.84$, $\mathfrak{D}=0.8$, and $\max_{\Lambda\times\lbrack 0,L\rbrack}\vert S(x,z)\vert^2=17.79$. (b): $\log_{10}U_{\rm max}=27.69$, $\mathfrak{D}=0.82$, and $\max_{\Lambda\times\lbrack 0,L\rbrack}\vert S(x,z)\vert^2=17.12$.}\label{figurelarged}
\end{figure}

We have compared the realizations in $\lbrace S\rbrace_{1\%}$ and $\lbrace S\rbrace\setminus\lbrace S\rbrace_{1\%}$ near the edge of the sampled domain of $U_{\rm max}$, where $\lbrace S\rbrace_{1\%}$ becomes statistically significant according to the results in Fig.~\ref{figureratio}. We have considered realizations with $\log_{10}U_{\rm max}\ge 27$. There are $15$ such realizations in $\lbrace S\rbrace$ among which $5$ in $\lbrace S\rbrace_{1\%}$ ($\mathfrak{D}\le 0.7$) and $10$ in $\lbrace S\rbrace\setminus\lbrace S\rbrace_{1\%}$ ($\mathfrak{D}> 0.7$). In Figures~\ref{figuresmalld} and \ref{figurelarged} we show two pairs of typical realizations picked in $\lbrace S\rbrace_{1\%}$ and  $\lbrace S\rbrace\setminus\lbrace S\rbrace_{1\%}$, respectively (technical details are given in the captions). For each realization, the theoretical instanton arriving at $x(L)=y_{\rm min}$ is indicated by a dashed contour, solution to $\vert S_{\rm inst}(x-y_{\rm min},z)\vert^2=0.75$ with $\vert S_{\rm inst}\vert^2$ as in Fig.~\ref{figureinst}. Intense localized hot spots similar to the theoretical one in Fig.~\ref{figureinst}(b) are clearly visible in both figures. In Fig.~\ref{figuresmalld} ($\mathfrak{D}\le 0.7$), hot spots occur inside the dashed line, in the instanton region. Note also that the level of $\vert S(x,z)\vert^2$ is significantly higher than average throughout the instanton region ($\sim 6$, while $\langle\vert S(x,z)\vert^2\rangle =1$), which seems difficult to explain by generic fluctuations (i.e. independent, small-scale hot spots). On the other hand, in Fig.~\ref{figurelarged} ($\mathfrak{D}> 0.7$), hot spots occur anywhere and the levels of $\vert S(x,z)\vert^2$ inside and outside the instanton region are quite comparable (hot spots excluded).

\begin{figure}[!h]
\includegraphics[width = 0.9\linewidth]{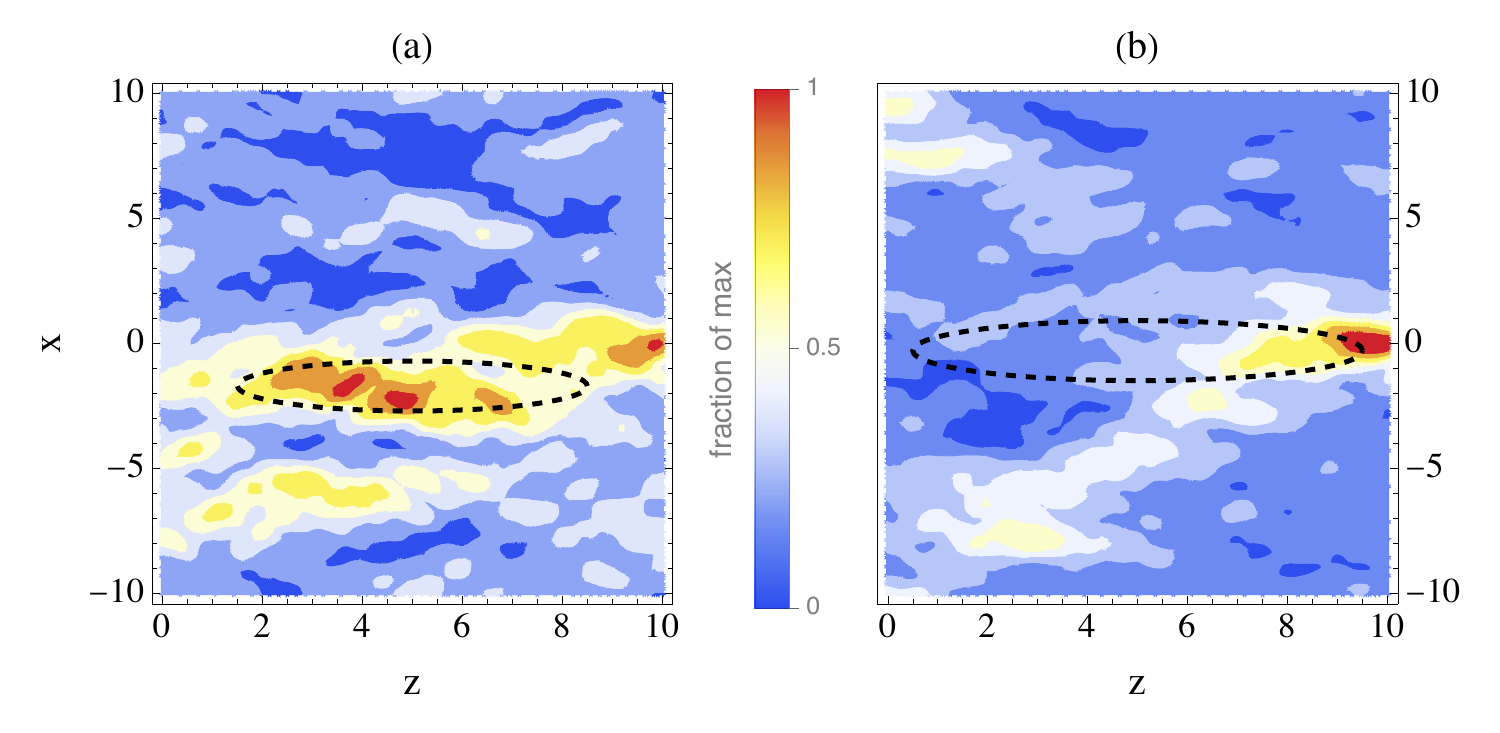}
\caption{Contour plots of the sample mean of $\vert\hat{S}(x+x_{\rm max},z)\vert^2$ for the realizations in (a) $\lbrace S\rbrace_{1\%}$ and (b) $\lbrace S\rbrace\setminus\lbrace S\rbrace_{1\%}$, with $\log_{10}U_{\rm max}\ge 27$. The dashed contours indicate the region where the sample mean of $\vert\hat{S}_{\rm inst}(x+x_{\rm max}-y_{\rm min},z)\vert^2$ is greater than $75\%$ of its maximum value.}\label{figureaverage}
\end{figure}

The robustness of these observations from one realization to the other can be tested through the sample mean of $\vert\hat{S}(x+x_{\rm max},z)\vert^2$ in which the realizations are translated to align the maxima of $\vert\psi(x,L)\vert^2$ with each other at the same position (here, $x=0$). In Figure~\ref{figureaverage}, we show the results for the same $15$ realizations with $\log_{10}U_{\rm max}\ge 27$ as above. The region of the sample mean of $\vert\hat{S}_{\rm inst}(x+x_{\rm max}-y_{\rm min},z)\vert^2$ is indicated by a dashed contour.

\begin{figure}[!h]
\includegraphics[width = 0.9\linewidth]{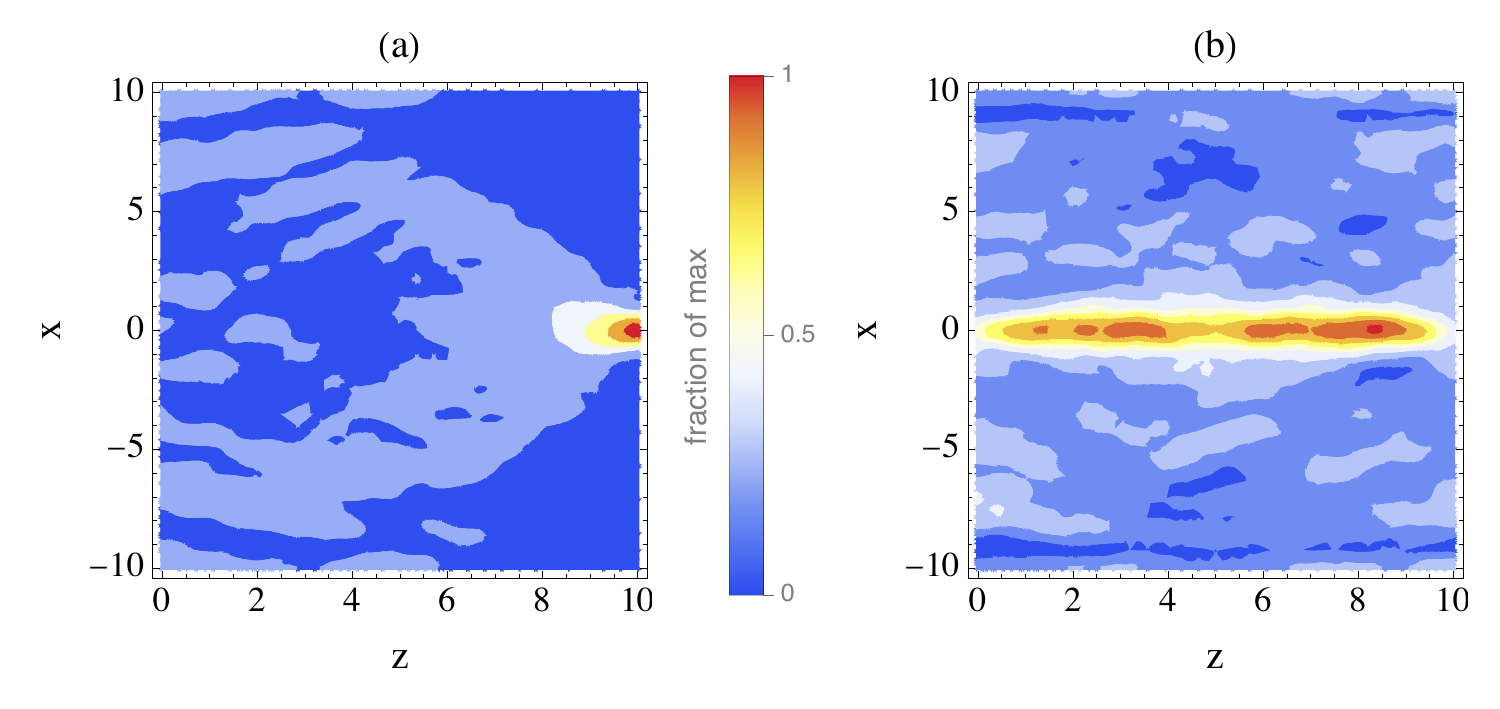}
\caption{Contour plots of the sample mean of $\vert\hat{S}(x+x_{\rm max},z)\vert^2$ for $200$ generic realizations of $S$ with $\log_{10}U_{\rm max}\le 10$ in the bulk of $p(\log_{10}U_{\rm max})$, (a) with diffraction ($m=0.7$) and (b) without diffraction ($m=+\infty$).}\label{figureaverage2}
\end{figure}

The overintensity in both figures (a) and (b) at $x=0$ and $z\simeq L$ is an effect of diffraction characteristic of the sub-asymptotic regime ($\log_{10}U_{\rm max}\not\gg 1$), as we will now explain. Consider first a generic realization of $S$ with $\log_{10}U_{\rm max}$ in the bulk of $p(\log_{10}U_{\rm max})$. In this case, $S$ is a collection of hot spots\ \cite{Dixit1993,G1985,RD1993} --- or a `hot spot field' --- with no visible instanton. Assume for the sake of argument that there is only one hot spot. For a given $U_{\rm max}$ the hot spot is either right behind the maximum of $\vert\psi (x,L)\vert^2$ (looking from $z>L$), or away from it with a larger amplitude to compensate for the diffraction loss. The most probable configuration is a compromise between those two options, and the fast decreasing probability of the hot spot intensity tilts the compromise in favor of the former. Namely, the hot spot is more likely to find itself right behind the maximum of $\vert\psi (x,L)\vert^2$. Generalizing to several hot spots, the same reasoning leads to the same conclusion: a generic realization of $S$ is more likely to have one hot spot right behind the maximum of $\vert\psi (x,L)\vert^2$ than not. This bias is cumulative in the calculation of the sample mean of $\vert\hat{S}(x+x_{\rm max},z)\vert^2$ which, as a result, is maximum at $x=0$ and $z\simeq L$. The situation is completely different if diffraction is switched off ($m^{-1}=0$). In this case, only the amplification along the straight path $x(z)=x_{\rm max}$ contributes to the maximum of $\vert\psi (x,L)\vert^2$, without diffraction loss, and the contributing hot spots can be anywhere along this path. This is exemplified in Figs.~\ref{figureaverage2}(a) and (b) that show the sample mean of $\vert\hat{S}(x+x_{\rm max},z)\vert^2$ for $200$ generic realizations of $S$ with $\log_{10}U_{\rm max}\le 10$ in the bulk of $p(\log_{10}U_{\rm max})$ (see Fig.~\ref{figureprobau}), with and without diffraction, respectively.

The results for $\log_{10}U_{\rm max}\ge 27$ in Fig.~\ref{figureaverage} are similar to those in Fig.~\ref{figureaverage2}(a) except that the instanton begins to emerge from the hot spot background in a non-negligible fraction of realizations ($\sim 30\%$), as seen in Figs.~\ref{figureratio}, \ref{figuresmalld} and \ref{figureaverage}(a). In conclusion, the overintensity in both Figs.~\ref{figureaverage}(a) and (b) at $x=0$ and $z\simeq L$ is an effect of the diffraction induced bias of the hot spot field, unrelated to the asymptotic instanton solutions. We refer the interested reader to the diffraction-free results for the sample mean of $\vert S\vert^2$ in\ \cite{MD2004} that show the emergence of the instanton but no overintensity near $z=L$, in agreement with our analysis. Lastly, the realizations of $S$ in the asymptotic limit $\ln U_{\rm max}\to +\infty$ are instanton dominated with negligible hot spot contribution to $\vert\psi (x,L)\vert^2$ near its maximum. So, in this limit, it makes no difference whether or not there is a hot spot right behind the maximum of $\vert\psi (x,L)\vert^2$ and there should be no diffraction induced bias of the (subdominant) hot spot field in the asymptotic regime.
\begin{figure}[ht]
\includegraphics[width = 0.8\linewidth]{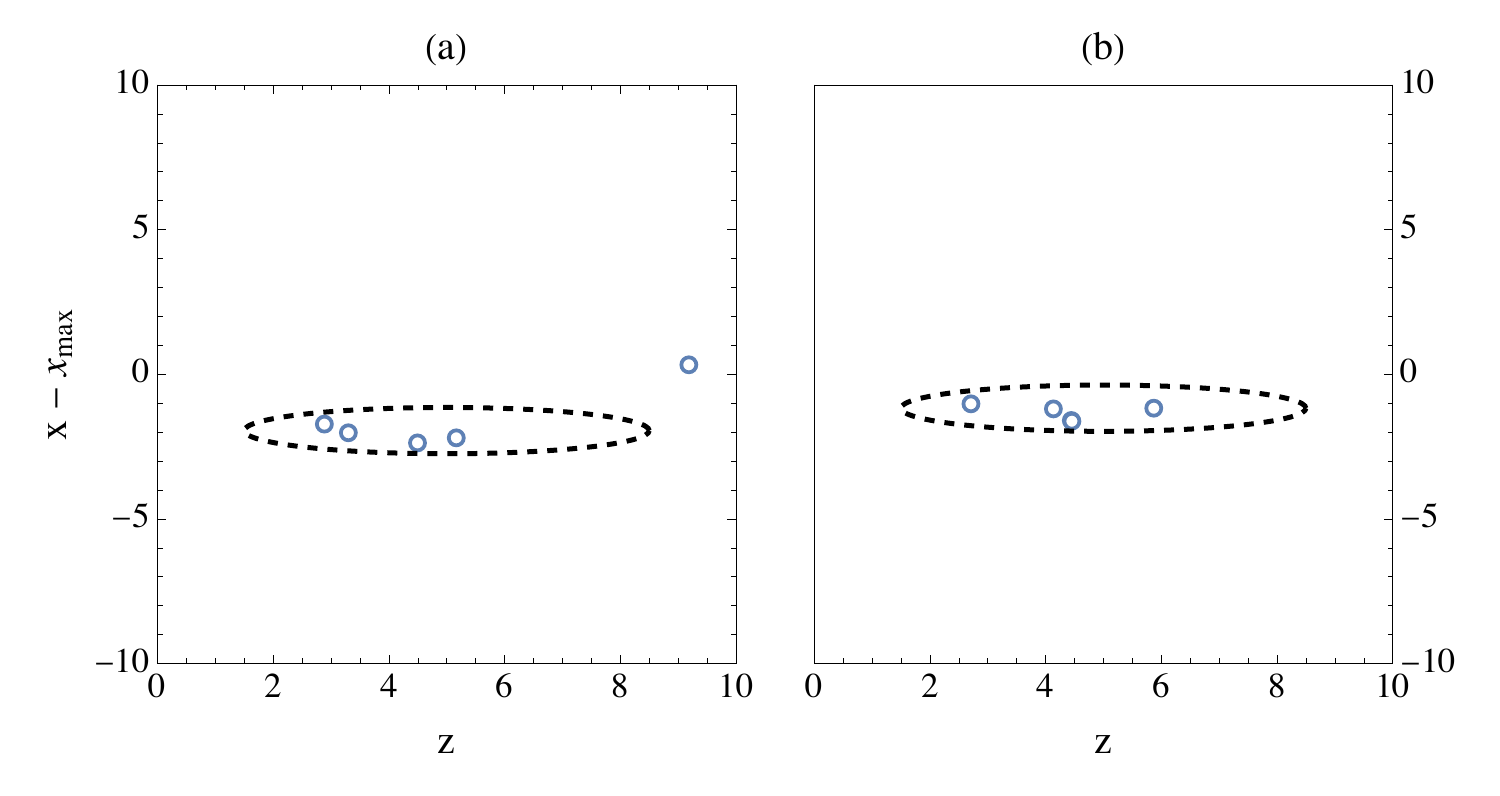}
\caption{Positions of the local maxima of $\vert S(x,z)\vert^2$ higher than $75\%$ of the global maximum (blue circles) for the same realizations as in Fig.~\ref{figuresmalld}. High maxima cluster in the instanton region indicated by the dashed contour. (See caption of Fig.~\ref{figuresmalld} for details.)}\label{figurecluster}
\end{figure}

We now turn to the rest of Fig.~\ref{figureaverage}. Figure~\ref{figureaverage}(a) shows the result for the realizations in $\lbrace S\rbrace_{1\%}$. In substance, it confirms the observations already made about the Fig.~\ref{figuresmalld}; namely, the observed level of $\vert S(x,z)\vert^2$ inside the dashed contour is the superposition of an average elevation of the level (the emerging instanton) and fluctuations of comparable amplitude. The presence of such an average elevation inside the dashed contour increases the probability that high maxima of $\vert S(x,z)\vert^2$ occur inside the instanton region. It is a pure statistical effect similar to the well known enhancement of correlations of peaks in Gaussian fields\ \cite{Kaiser1984,BBKS1986}, the large-scale instanton playing the same role as the `signal' and `background field' in\ \cite{Kaiser1984} and\ \cite{BBKS1986}, respectively. As a consequence, one observes (i) a tendency for the hot spots to cluster in the instanton region and (ii) a level of $\vert S(x,z)\vert^2$ between the hot spots significantly higher than the average level outside the instanton region. These two points (i) and (ii) signal the emergence of the instanton in the realizations of $\lbrace S\rbrace_{1\%}$. Hot spot clustering is illustrated in Fig.~\ref{figurecluster} for the same realizations as in Fig.~\ref{figuresmalld}. By contrast, no particular structure is observed in Fig.~\ref{figureaverage}(b) for the realizations in $\lbrace S\rbrace\setminus\lbrace S\rbrace_{1\%}$ (except the overintensity at $x=0$ and $z\simeq L$). It means that neither emerging instanton nor clustering of hot spots are significant in those realizations.

Combining numerical results with analytical predictions, we can now infer how the transition to the asymptotic regime occurs as $\ln U$ increases. As long as the value of $\ln U$ is in the bulk of $p(\ln U)=Up(U)$, the overwhelming majority of the realizations of $S$ are generic realizations with hot spots uniformly scattered in $\Lambda\times\lbrack 0,L\rbrack$ and $\mathfrak{D}$ close to its typical value at $\mathfrak{D}\simeq 0.86$. The situation changes gradually as $\ln U$ increases into the tail of $p(\ln U)$, as seen in Fig.~\ref{figureratio}. Namely, the larger $\ln U$ the larger the percentage of atypical realizations with $\mathfrak{D}$ smaller than, say, its first percentile --- like the ones in Fig.~\ref{figuresmalld} --- to the detriment of generic realizations --- like the ones in Fig.~\ref{figurelarged}. In those atypical realizations, the hot spots cluster in the instanton region instead of being uniformly scattered in $\Lambda\times\lbrack 0,L\rbrack$ and the level of $\vert S(x,z)\vert^2$ between the hot spots remains abnormally high (see Figs.~\ref{figuresmalld}, \ref{figureaverage}(a), and \ref{figurecluster}). Letting $\ln U\to +\infty$, the percentage of atypical realizations goes up to $100\%$ while $\mathfrak{D}$ and the relative fluctuations-to-instanton amplitude decrease to zero with probability one. In this limit, the tail of $p(U)$ is asymptotically dominated by the instanton which determines the the critical coupling $g_c(L)$.
%
%
\section{Discussion and perspectives}\label{conclusion}
In this paper, we have studied the large amplification limit of a linear amplifier driven by the square of a Gaussian random field. We have considered the same model as in Refs.~\cite{RD1994} and\ \cite{MCL2006} in which the propagation is that of a free Schr\"{o}dinger equation. By performing the first instanton analysis of the corresponding MSR action, we have identified the realizations of the Gaussian field most likely to produce a large amplification. We have found that when $\ln U=\ln\vert\psi(0,L)\vert^2$ gets large, for $\psi$ solution to Eq.~(\ref{withDeq}) with $S$ defined in Sec.\ \ref{modelanddef}, the realizations of $S$ concentrate onto large-scale filamentary instantons running along the path(s) maximizing the largest eigenvalue of the covariance operator defined in Eq.~(\ref{covariance3Dpath}). This result explains the otherwise mysterious presence of this maximized eigenvalue in the expression of $g_c(L)$ found in\ \cite{MCL2006} (see Eq.~(\ref{critcouplingwithD})). We have then derived the tail of $p(U)$ for large $\ln U$ from the instanton contribution and checked that the resulting critical coupling does coincide with the one in Ref.\ \cite{MCL2006}. From this analysis, it follows in particular that the realizations of $S$ causing the divergence of $\langle\vert\psi(0,L)\vert^2\rangle$ for $g>g_c(L)$ are long filamentary structures (the instantons) rather than localized hot-spots, as assumed in hot-spot models~\cite{RD1994}. This result extends the conclusions of Ref.~\cite{MD2004} to the full problem~(\ref{withDeq}) with diffraction.

Numerical simulations clearly show a statistical bias of $S$ towards the instanton, as $\ln U$ increases. The larger $\ln U$ in the sampled range, the larger the fraction of atypical realizations of $S$ in which a large-scale instanton coexists with fluctuation induced localized hot spots. (See\ \cite{MD2004} for a quantitative comparison of hot spot and instanton contributions to the amplification in the diffraction-free case.) In those atypical realizations, hot spots are not uniformly distributed in $\Lambda\times\lbrack 0,L\rbrack$ but tend to cluster in the instanton region. For the experimentally realistic sample size we considered ($\sim 10^5$), it proved impossible to probe values of $\ln U$ large enough that the fluctuations of $S$ away from the instanton could be neglected. Hot spot clustering and nonlinear evolution of the coupled hot spots/instanton system are interesting subjects that would deserve to be dealt with in more depth, especially in laser-plasma interaction physics.

The work presented here is only a first step toward a comprehensive study of Eq.~(\ref{withDeq}) in the large amplification limit. There are various directions along which investigations could be pushed further. Obviously, trying to lift all or part of the assumptions made in Secs.~\ref{modelanddef} and\ \ref{singlepathsec} appears as a natural next step, especially the technical restriction (ii) and the possibility of multi-filament instantons. As mentioned at the beginning of Sec.~\ref{numerics}, it would also be important to directly test the validity of our analytical results, either by significantly increasing the sample size, or by using a biased numerical scheme capable of probing $p(U)$ in the asymptotic regime.

Another challenging line of research is the study of a possible intermittency of $\vert\psi(x,L)\vert^2$ and its connection with our results, as we will now explain. The experimental conditions to which our results can be directly applied are those that naturally sample the realizations of $S$, like, e.g., in a laser-plasma interaction experiment with space-time optical smoothing in which $S$ is renewed periodically. The situation is different in the case of purely spatial smoothing, where a unique realization of $S$ is available in a given experimental environment and $\langle\vert\psi(0,L)\vert^2\rangle$ is replaced with the space average $\vert\Lambda\vert^{-1}\int_{\Lambda}\vert\psi(x,L)\vert^2\, dx$ for a generic realization of $S$. As rare events, instantons are very unlikely to contribute to the latter quantity unless $\vert\Lambda\vert$ is large and the space average is dominated by the contribution of scarce, intense peaks of $\vert\psi(x,L)\vert^2$ the high amplitude of which outbalances their scarcity. The question is then whether such a peak-dominated behavior --- called `intermittency' in the literature on random media \cite{Molchanov1991} --- can be observed in the solution to Eq.~(\ref{withDeq}) for large $\vert\Lambda\vert$ and $g>g_c(L)$. If so, our results imply that $S(x,z)$ in the region of $\Lambda\times\lbrack 0,L\rbrack$ upstream from a dominant peak of $\vert\psi(x,L)\vert^2$ is a filament instanton arriving at the peak location. Intermittency of $\vert\psi(x,L)\vert^2$ is thus important as connecting our instanton analysis approach with experimental results for a given realization of $S$ in the large $\vert\Lambda\vert$ limit and $g>g_c(L)$. The interested reader will find a detailed introduction to intermittency in random media in Ref.~\cite{Molchanov1991}.

Finally, it would also be interesting to investigate the small $m$ behavior of the same problem. For $m\to 0$, it can be shown that $\psi(x,z)$ reduces to
\begin{equation}\label{m-to-zero}
\psi(x,z)=\exp\left(\frac{g}{\vert\Lambda\vert}\, \| S\|^2_{2,\Lambda\times\lbrack 0,L\rbrack}\right) .
\end{equation}
Thus, in this limit, the realizations of $S$ giving rise to a large $\ln U$ are the ones with a large $L^2$-norm, which are known to concentrate onto the fundamental eigenspace of the covariance operator $T_C$ defined in Eq.\ (\ref{covariance3D1}), as $\ln U\to +\infty$\ \cite{MD2004,MC2011}. If $S$ is given by the random Fourier sum~(\ref{KLexpansion}) with, e.g., $\sigma_{(n,j)}<\sigma_{(0,0)}$ for all $(n,j)\ne (0,0)$, the fundamental eigenspace of $T_C$ reduces to the functions $\propto\Phi_{(0,0)}(z)$ independent of $x$, and the realizations of $S$ in the large $\ln U$ limit are completely delocalized in $\Lambda$, in striking contrast to the filamentary instantons we have found for a fixed $m\ne 0$. This simple example indicates that the two limits $\ln U\to +\infty$ and $m\to 0$ do not commute, which raises the natural question of how precisely the crossover between `$\ln U\to +\infty$ then $m\to 0$' and \mbox{`$m\to 0$ then $\ln U\to +\infty$'} occurs. Answering this question will elucidate the intriguing transition suggested by the above example, from filamentary to delocalized instantons, as $m$ goes to zero.

In conclusion, it may be noted that the number of highly non-trivial questions raised by the seemingly simple linear problem (\ref{withDeq}) is quite remarkable. Following on from the work presented here, we hope that those questions will motivate interesting research in both statistical physics and laser-matter interaction physics where the linear amplifier model (\ref{withDeq}) first appeared.
%
%
\acknowledgments{The author warmly thanks Satya N Majumdar, Denis Pesme, and Gr\'egory Schehr for their interest and valuable advice about the manuscript. He also thanks Harvey A Rose and Joel L Lebowitz for the inspiring discussions he had with them on related subjects.}
%
%
\appendix
\section{Paths maximizing $\bm{\mu_1\lbrack x(\cdot)\rbrack}$ and ridge paths of $\bm{\vert S_{\rm inst}(x,z)\vert^2}$}\label{app1}
In this appendix we show that $x_{\rm inst}(\cdot)$ is a ridge path of $\vert S_{\rm inst}^{x_{\rm inst}(\cdot)}(x,z)\vert^2$. From Eqs.~(\ref{matrixM}) and (\ref{instsolwithDS1}), one gets
\begin{equation}\label{app1eq2}
\int_0^L\vert S_{\rm inst}^{x_{\rm inst}(\cdot)}(x_{\rm inst}(z),z)\vert^2\, dz=
\bm{\mathfrak{s}}^\dagger M\lbrack x_{\rm inst}(\cdot)\rbrack\bm{\mathfrak{s}}
=\mu_{\rm max}\|\bm{\mathfrak{s}}\|^2 ,
\end{equation}
where $\bm{\mathfrak{s}}$ is in the fundamental eigenspace of $M\lbrack x_{\rm inst}(\cdot)\rbrack$, and
\begin{eqnarray}\label{app1eq3}
&&\int_0^L\vert S_{\rm inst}^{x_{\rm inst}(\cdot)}(x(z),z)\vert^2\, dz =
\bm{\mathfrak{s}}^\dagger M\lbrack x(\cdot)\rbrack\bm{\mathfrak{s}}
\le\mu_1\lbrack x(\cdot)\rbrack\|\bm{\mathfrak{s}}\|^2 \nonumber \\
&&=\frac{\mu_1\lbrack x(\cdot)\rbrack}{\mu_{\rm max}}
\int_0^L\vert S_{\rm inst}^{x_{\rm inst}(\cdot)}(x_{\rm inst}(z),z)\vert^2\, dz
\le\int_0^L\vert S_{\rm inst}^{x_{\rm inst}(\cdot)}(x_{\rm inst}(z),z)\vert^2\, dz,
\end{eqnarray}
yielding
\begin{equation}\label{app1eq4}
\sup_{x(\cdot)\in B(0,L)}\int_0^L\vert S_{\rm inst}^{x_{\rm inst}(\cdot)}(x(z),z)\vert^2\, dz =
\int_0^L\vert S_{\rm inst}^{x_{\rm inst}(\cdot)}(x_{\rm inst}(z),z)\vert^2\, dz.
\end{equation}
Equation\ (\ref{app1eq4}) means that in the path-integral for $\varphi_{\rm inst}^{x_{\rm inst}(\cdot)}(0,L)$, $x_{\rm inst}(\cdot)$ is a path along which the amplification is maximum. Now, assume that there is $A\subset\lbrack 0,L\rbrack$ with $\vert A\vert\equiv\int_0^L\bm{1}_{z\in A} dz>0$ such that for all $z\in A$, there is $x\in\Lambda$ with $\vert S_{\rm inst}^{x_{\rm inst}(\cdot)}(x,z)\vert^2>\vert S_{\rm inst}^{x_{\rm inst}(\cdot)}(x_{\rm inst}(z),z)\vert^2$. It follows immediately that
\begin{equation}\label{app1eq5}
\int_0^L \sup_{x\in\Lambda}\vert S_{\rm inst}^{x_{\rm inst}(\cdot)}(x,z)\vert^2 dz >
\int_0^L\vert S_{\rm inst}^{x_{\rm inst}(\cdot)}(x_{\rm inst}(z),z)\vert^2\, dz,
\end{equation}
and from Eqs.~(\ref{app1eq4}) and~(\ref{app1eq5}) one should have
\begin{equation}\label{app1eq6}
\int_0^L \sup_{x\in\Lambda}\vert S_{\rm inst}^{x_{\rm inst}(\cdot)}(x,z)\vert^2 dz >
\sup_{x(\cdot)\in B(0,L)}\int_0^L\vert S_{\rm inst}^{x_{\rm inst}(\cdot)}(x(z),z)\vert^2\, dz,
\end{equation}
in contradiction with the Lemma A1 in Ref.\ \cite{MCL2006} according to which one must have an equality. Thus, there is no such $A$ and since every given realization of $S_{\rm inst}^{x_{\rm inst}(\cdot)}(x,z)$ in Eq.~(\ref{instsolwithDS1}) is a continuous function of $x$ and $z$, one has $\vert S_{\rm inst}^{x_{\rm inst}(\cdot)}(x,z)\vert^2\le\vert S_{\rm inst}^{x_{\rm inst}(\cdot)}(x_{\rm inst}(z),z)\vert^2$ for all $0\le z\le L$ and $x\in\Lambda$. This proves that for all the realizations of $S_{\rm inst}^{x_{\rm inst}(\cdot)}(x,z)$ in Eq.~(\ref{instsolwithDS1}) with $\bm{\mathfrak{s}}$ in the fundamental eigenspace of $M\lbrack x_{\rm inst}(\cdot)\rbrack$, $x_{\rm inst}(\cdot)$ is a ridge path of $\vert S_{\rm inst}^{x_{\rm inst}(\cdot)}(x,z)\vert^2$ along which the amplification is maximum.

Assume that there is a $\bm{\mathfrak{s}}$ in the fundamental eigenspace of $M\lbrack x_{\rm inst}(\cdot)\rbrack$ and $y_{\rm inst}(\cdot)\in B(0,L)$ with $y_{\rm inst}(\cdot)\ne x_{\rm inst}(\cdot)$ such that $y_{\rm inst}(\cdot)$ is also a ridge path of $\vert S_{\rm inst}^{x_{\rm inst}(\cdot)}(x,z)\vert^2$ along which the amplification is maximum. Then, $y_{\rm inst}(\cdot)$ maximizes $\mu_1\lbrack x(\cdot)\rbrack$ and $\bm{\mathfrak{s}}$ belongs to the fundamental eigenspace of $M\lbrack y_{\rm inst}(\cdot)\rbrack$ (otherwise, the amplification along $y_{\rm inst}(\cdot)$ would be less than along $x_{\rm inst}(\cdot)$). Since $\bm{\mathfrak{s}}$ belongs to the fundamental eigenspaces of both $M\lbrack x_{\rm inst}(\cdot)\rbrack$ and $M\lbrack y_{\rm inst}(\cdot)\rbrack$, their intersection is necessarily non trivial. It shows that the number of ridge paths depends on the relative structure of the fundamental eigenspaces of $M\lbrack x(\cdot)\rbrack$ for the different paths maximizing $\mu_1\lbrack x(\cdot)\rbrack$. If the fundamental eigenspaces of $M\lbrack x(\cdot)\rbrack$ for all the paths maximizing $\mu_1\lbrack x(\cdot)\rbrack$ are essentially disjoint, $\bm{\mathfrak{s}}$ cannot belong to more than one fundamental eigenspace and each realization of the instanton has only one ridge path. This is the case considered in the paper. On the other hand, if the fundamental eigenspaces of $M\lbrack x(\cdot)\rbrack$ for different paths maximizing $\mu_1\lbrack x(\cdot)\rbrack$ have a non trivial intersection, then for all the realizations with $\bm{\mathfrak{s}}$ in the intersection, the instanton has more than one ridge path. This case corresponds to multi-filament instantons.
%
%
\section{Equivalence of the Fourier and convolution representations of $\bm{S_{\rm inst}}$}\label{app2}
In this appendix, we prove the equivalence of the expressions of $S_{\rm inst}^{x_{\rm inst}(\cdot)}(x,z)$ in Eqs.~(\ref{instsolwithDS1}) and (\ref{instsolwithDS2}). Permuting the sum and the integral on the right-hand side of Eq.~(\ref{instsolwithDS2}) and using the Fourier decomposition\ (\ref{CFexpansion}) for $C(x-x_{\rm inst}(z^\prime),z,z^\prime)$, one readily finds that the equation (\ref{instsolwithDS2}) can be rewritten as
\begin{equation}\label{instsolwithDS3}
S_{\rm inst}^{x_{\rm inst}(\cdot)}(x,z)= c_1 \Omega_1(x,z),
\end{equation}
with
\begin{equation}\label{Omegafunction}
\Omega_1(x,z)=\sum_{(n,j)\in\mathcal{I}}
\mathfrak{e}^{(1)}_{(n,j)}
\sqrt{\frac{\sigma_{(n,j)}}{\ell\, \mu_{\rm max}}}\, 
{\rm e}^{2i\pi nx/\ell}\Phi_{(n,j)}(z),
\end{equation}
where $\bm{\mathfrak{e}}^{(1)}$ is a vector defined by its components
\begin{equation}\label{vector-e_app2}
\mathfrak{e}^{(1)}_{(n,j)}=
\sqrt{\frac{\sigma_{(n,j)}}{\ell\, \mu_{\rm max}}}\, 
\int_0^L{\rm e}^{-2i\pi nx_{\rm inst}(z^\prime)/\ell}\Phi_{(n,j)}(z^\prime)^\ast
\phi_1(z^\prime)\, dz^\prime .
\end{equation}
Showing that the equation~(\ref{instsolwithDS1}) can also be written in the form of Eq.~(\ref{instsolwithDS3}) requires a little more work. From Eqs.~(\ref{CFexpansion}), (\ref{matrixM}), and (\ref{vector-e_app2}) it can be checked that
\begin{eqnarray}\label{Mapplyto-e}
&&\left(M\lbrack x_{\rm inst}(\cdot)\rbrack \bm{\mathfrak{e}}^{(1)}\right)_{(n,j)} =
\sum_{(m,k)\in\mathcal{I}}M_{(n,j)(m,k)}\lbrack x_{\rm inst}(\cdot)\rbrack
\, \mathfrak{e}^{(1)}_{(m,k)} \nonumber \\
&&=\sqrt{\frac{\sigma_{(n,j)}}{\ell\, \mu_{\rm max}}}\, 
\int_0^L{\rm e}^{-2i\pi nx_{\rm inst}(z)/\ell}\Phi_{(n,j)}(z)^\ast
\, \langle z\vert T_{x_{\rm inst}(\cdot)}\vert \phi_1\rangle\, dz \nonumber \\
&&= \mu_{\rm max}\, \mathfrak{e}^{(1)}_{(n,j)},
\end{eqnarray}
and
\begin{eqnarray}\label{scalarproduct-e}
&&\|\bm{\mathfrak{e}}^{(1)}\|^2=
\sum_{(n,j)\in\mathcal{I}}\vert\mathfrak{e}^{(1)}_{(n,j)}\vert^2
=\frac{1}{\mu_{\rm max}}
\, \langle\phi_1\vert T_{x_{\rm inst}(\cdot)}\vert \phi_1\rangle \nonumber \\
&&=\langle\phi_1\vert\phi_1\rangle =1,
\end{eqnarray}
which means that $\bm{\mathfrak{e}}^{(1)}$ is the normalized fundamental eigenvector of $M\lbrack x_{\rm inst}(\cdot)\rbrack$. Writing
\begin{equation}\label{stoc-ctos_app2}
\mathfrak{s}_{(n,j)}=\frac{1}{\sqrt{\mu_{\rm max}}}
\, c_1 \mathfrak{e}^{(1)}_{(n,j)}
\ {\rm with}\ 
c_1=\sqrt{\mu_{\rm max}}
\, \sum_{(n,j)\in\mathcal{I}}\mathfrak{s}_{(n,j)}\mathfrak{e}^{(1)\, \ast}_{(n,j)},
\end{equation}
on the right-hand side of Eq.~(\ref{instsolwithDS1}), one obtains the same equation~(\ref{instsolwithDS3}), as expected, which proves the equivalence of Eqs.~(\ref{instsolwithDS1}) and (\ref{instsolwithDS2}) with $\mathfrak{s}_{(n,j)}$ and $c_1$ related to each other by Eq.~(\ref{stoc-ctos_app2}).
%
%
\section{Limit of $\bm{\eta^{-1}\ln A(\eta)}$ and $\bm{\partial_{\eta}\ln A(\eta)}$ as $\bm{\eta\to +\infty}$}\label{app3}
In this appendix we derive the two limits in Eq.\ (\ref{limitsofA}). We will use the convolution representation\ (\ref{instsolwithDS2}). To make the dependence of $S_{\rm inst}^{x_{\rm inst}(\cdot)}$ on $c_1$ explicit we write $S_{\rm inst}^{x_{\rm inst}(\cdot)}(x,z)\equiv S_{\rm inst}^{x_{\rm inst}(\cdot)}(x,z,c_1)=\sqrt{\eta}\, {\rm e}^{i\arg(c_1)}\, S_{\rm inst}^{x_{\rm inst}(\cdot)}(x,z,1)$, where $\eta=\vert c_1\vert^2$. Deriving the Feynman-Kac path-integral representation of $\varphi_{\rm inst}^{x_{\rm inst}(\cdot)}(0,L)$,
\begin{equation}\label{app3eq1}
\varphi_{\rm inst}^{x_{\rm inst}(\cdot)}(0,L)=
\int_{x(\cdot)\in B(0,L)}{\rm e}^{\int_{0}^{L}
\left\lbrack\frac{im}{2}\dot{x}(\tau)^2+g\eta\vert S_{\rm inst}^{x_{\rm inst}(\cdot)}(x(\tau),\tau,1)\vert^2\right\rbrack\, d\tau}
\mathscr{D}x,
\end{equation}
with respect to $\eta$ and using the fact that $\int_{0}^{L}
\vert S_{\rm inst}^{x_{\rm inst}(\cdot)}(x_{\rm inst}(\tau),\tau,1)\vert^2 d\tau =1$, one gets
\begin{eqnarray}\label{app3eq2}
\partial_\eta\varphi_{\rm inst}^{x_{\rm inst}(\cdot)}(0,L)&=&g\int_{x(\cdot)\in B(0,L)}
\left(\int_{0}^{L}\vert S_{\rm inst}^{x_{\rm inst}(\cdot)}(x(\tau),\tau,1)\vert^2 d\tau\right) \nonumber \\
&&\times\, {\rm e}^{\int_{0}^{L}
\left\lbrack\frac{im}{2}\dot{x}(\tau)^2+g\eta\vert S_{\rm inst}^{x_{\rm inst}(\cdot)}(x(\tau),\tau,1)\vert^2\right\rbrack\, d\tau}
\mathscr{D}x \nonumber \\
&\sim& g\int_{x(\cdot)\in B(0,L)}\left(\int_{0}^{L}
\vert S_{\rm inst}^{x_{\rm inst}(\cdot)}(x_{\rm inst}(\tau),\tau,1)\vert^2 d\tau\right) \nonumber \\
&&\times\, {\rm e}^{\int_{0}^{L}
\left\lbrack\frac{im}{2}\dot{x}(\tau)^2+g\eta\vert S_{\rm inst}^{x_{\rm inst}(\cdot)}(x(\tau),\tau,1)\vert^2\right\rbrack\, d\tau}
\mathscr{D}x \nonumber \\
&=&g\int_{x(\cdot)\in B(0,L)}{\rm e}^{\int_{0}^{L}
\left\lbrack\frac{im}{2}\dot{x}(\tau)^2+g\eta\vert S_{\rm inst}^{x_{\rm inst}(\cdot)}(x(\tau),\tau,1)\vert^2\right\rbrack\, d\tau}
\mathscr{D}x \nonumber \\
&=&g\varphi_{\rm inst}^{x_{\rm inst}(\cdot)}(0,L)\ \ \ \ \ (\eta\to +\infty),
\end{eqnarray}
from which it follows that
\begin{equation}\label{app3eq3}
\partial_\eta\ln\vert\varphi_{\rm inst}^{x_{\rm inst}(\cdot)}(0,L)\vert^2
=2{\rm Re}\left(\frac{\partial_\eta\varphi_{\rm inst}^{x_{\rm inst}(\cdot)}(0,L)}
{\varphi_{\rm inst}^{x_{\rm inst}(\cdot)}(0,L)}\right)
\sim 2g\ \ \ \ \ (\eta\to +\infty).
\end{equation}
Thus, for all $\varepsilon >0$ there is $\eta_0>0$ such that for every $\eta\ge\eta_0$,
\begin{equation}\label{app3eq4}
2g(1-\varepsilon)\le
\partial_\eta\ln\vert\varphi_{\rm inst}^{x_{\rm inst}(\cdot)}(0,L)\vert^2
\le 2g(1+\varepsilon).
\end{equation}
Writing $\vert\varphi_{\rm inst}^{x_{\rm inst}(\cdot)}(0,L)\vert^2=A(\eta)\, {\rm e}^{2g\eta}$ in Eq.~(\ref{app3eq4}), one obtains
\begin{equation}\label{app3eq5}
-2g\varepsilon\le\partial_\eta\ln A(\eta)\le 2g\varepsilon ,
\end{equation}
for every $\eta\ge\eta_0$, and since $\varepsilon$ can be taken arbitrarily small, Eq.~(\ref{app3eq5}) reduces to
\begin{equation}\label{app3eq6}
\lim_{\eta\to +\infty}\partial_\eta\ln A(\eta)=0,
\end{equation}
which is the second limit in Eq.~(\ref{limitsofA}). To get the first limit, we integrate Eq.~(\ref{app3eq5}) from $\eta_0$ to any $\eta>\eta_0$, which yields
\begin{equation}\label{app3eq7}
-2g\varepsilon\eta +K_{+}(\varepsilon)
\le\ln A(\eta)\le
2g\varepsilon\eta +K_{-}(\varepsilon),
\end{equation}
with $K_{\pm}(\varepsilon)=\ln A(\eta_0)\pm 2g\varepsilon\eta_0$. Note that $\ln A(\eta_0)$ exists, otherwise $\ln A(\eta)$ would have a vertical asymptote at $\eta =\eta_0$, in contradiction with Eq.~(\ref{app3eq5}). It remains to divide Eq.~(\ref{app3eq7}) by $\eta$:
\begin{equation}\label{app3eq8}
-2g\varepsilon -\frac{\vert K_{+}(\varepsilon)\vert}{\eta}
\le\frac{1}{\eta}\ln A(\eta)\le
2g\varepsilon +\frac{\vert K_{-}(\varepsilon)\vert}{\eta},
\end{equation}
where we have used $K_{+}(\varepsilon)\ge -\vert K_{+}(\varepsilon)\vert$ and $K_{-}(\varepsilon)\le \vert K_{-}(\varepsilon)\vert$. Now, for $\eta$ large enough, namely $\eta\ge\max\lbrack\eta_0,\varepsilon^{-1}\vert K_{+}(\varepsilon)\vert,\varepsilon^{-1}\vert K_{-}(\varepsilon)\vert\rbrack$, Eq.~(\ref{app3eq8}) gives
\begin{equation}\label{app3eq9}
-3g\varepsilon\le\frac{1}{\eta}\ln A(\eta)\le 3g\varepsilon,
\end{equation}
and since $\varepsilon$ can be taken arbitrarily small, one finally obtains
\begin{equation}\label{app3eq10}
\lim_{\eta\to +\infty}\frac{1}{\eta}\ln A(\eta)=0,
\end{equation}
which is the first limit in Eq.~(\ref{limitsofA}).

We now prove the relation $\ln A((2g)^{-1}\ln U)^{1/2\mu_{\rm max}g} =o(\ln U)$ used in Eq.~(\ref{tailpofUwithD4}). For all $\varepsilon >0$ there is $\eta_0>0$ such that for every $\eta\ge\eta_0$,
\begin{equation}\label{app3eq11}
\exp(-\varepsilon\eta)\le A(\eta)\le\exp(\varepsilon\eta),
\end{equation}
whence,
\begin{equation}\label{app3eq12}
\exp\left(-\frac{\varepsilon\eta}{2\mu_{\rm max}g}\right)
\le A(\eta)^{1/2\mu_{\rm max}g}\le
\exp\left(\frac{\varepsilon\eta}{2\mu_{\rm max}g}\right),
\end{equation}
and, taking the logarithm,
\begin{equation}\label{app3eq13}
-\frac{\varepsilon\eta}{2\mu_{\rm max}g}
\le\ln A(\eta)^{1/2\mu_{\rm max}g}\le
\frac{\varepsilon\eta}{2\mu_{\rm max}g},
\end{equation}
for every $\eta\ge\eta_0$. Since $\varepsilon$ can be taken arbitrarily small, Eq.~(\ref{app3eq13}) reduces to
\begin{equation}\label{app3eq14}
\lim_{\eta\to +\infty}\frac{1}{\eta}
\ln A(\eta)^{1/2\mu_{\rm max}g}=0,
\end{equation}
which means that $\ln A(\eta)^{1/2\mu_{\rm max}g}=o(\eta)$, as announced.
%
%

%
\end{document}